\documentclass[aps,prb,twocolumn,superscriptaddress,longbibliography,10pt]{revtex4-2}
\usepackage{graphicx}
\usepackage{hyperref}
\usepackage{amssymb}
\usepackage{dcolumn}
\usepackage{float}
\usepackage{booktabs}
\usepackage{amsmath} 
\usepackage[utf8]{inputenc}
\usepackage{lmodern}
\usepackage{color}
\definecolor{red}{rgb}{1.0,0.0,0.0}
\usepackage{makecell}
\usepackage{marginnote}

\usepackage{natbib}
\usepackage{float}
\usepackage{placeins}

\newcommand{\e}{\mathrm{e}}

\DeclareMathAlphabet{\bi}{OML}{cmm}{b}{it}

\def\ba{\begin{aligned}}
	\def\ea{\end{aligned}}
\def\be{\begin{equation}}
	\def\ee{\end{equation}}
\def\bearr{\begin{eqnarray}}
	\def\eearr{\end{eqnarray}}

\def\l{\left}
\def\r{\right}

\usepackage{tikz}

\usepackage{xcolor}
\pagecolor{white}

\usepackage{xcolor}
\color{black}

\hypersetup{
	colorlinks=true,
	linkcolor=blue,
	citecolor=blue,
	urlcolor=blue
}

\begin{document}
	\title{Role of anisotropic confining potential and elliptical driving in dynamics of a Ge hole qubit}
	\bigskip
	
    \author{Bashab Dey}
    \email{Bashab.Dey@ur.de}
    \author{John Schliemann}
		\affiliation {Institute of Theoretical Physics, University of Regensburg, Regensburg, Germany}

	\begin{abstract}
The squeezing of a Ge planar quantum dot enhances the Rabi frequency of electric dipole spin resonance by several orders of magnitude due to a strong Direct Rashba spin-orbit interaction (DR-SOI) in such geometries \href{https://doi.org/10.1103/PhysRevB.104.115425} {\color{blue}[Phys. Rev. B {\bf 104}, 115425 (2021)]}. We investigate the geometric effect of an elliptical (squeezed) confinement  and its interplay with the polarization of driving field in determining the Rabi frequency of a heavy-hole qubit in a planar Ge quantum dot. 
To calculate the Rabi frequency, we consider only the $p$-linear SOIs viz. electron-like Rashba, hole-like Rashba and hole-like Dresselhaus which are claimed to be the dominant ones by recent studies on planar Ge heterostructures. We derive approximate analytical expressions of the Rabi frequency using a Schrieffer-Wolff transformation for small SOI and driving strengths. Firstly, for an out-of-plane magnetic field with magnitude $B$, we get an operating region with respect to $B$, squeezing and polarization parameters 
where the qubit can be operated to obtain `clean' Rabi flips. On and close to the boundaries of the region, the higher orbital levels strongly interfere with the two-level qubit subspace and destroy the Rabi oscillations, thereby putting a limitation on squeezing of the confinement. The Rabi frequency shows different behaviour for electron-like and hole-like Rashba SOIs. It vanishes for right (left) circular polarization in presence of purely electron-like (hole-like) Rashba SOI in a circular confinement. 
For both in- and out-of-plane magnetic fields, higher Rabi frequencies are achieved for squeezed configurations when the ellipses of polarization and the confinement equipotential have their major axes aligned but with different eccentricities. We also deduce a simple formula to calculate the effective heavy hole mass by measuring the Rabi frequencies using this setup.
 \end{abstract}

 \maketitle

\section{Introduction}

Hole spin qubits  have drawn immense interest in recent times due to several advantageous features over their electronic counterparts such as stronger spin-orbit interaction (SOI) enabling faster electrical manipulation \cite{edsr-loss-hole}, reduced contact-hyperfine interaction leading to longer decoherence times \cite{hole-hyperfine1,hole-hyperfine2,hole-hyperfine3,hole-hyperfine4}, and absence of valley degeneracy \cite{valley}.  These qubits are based on the valence band states of group IV (Si, Ge) and III-V (GaAs, InSb etc.) semiconductors \cite{hole-review}. Among them, Germanium turns out to be a favorite due to the low effective mass of holes \cite{hole-mass} which allows  larger dot sizes, isotropic purification \cite{Si-isotope, Si-isotope2,Ge-isotope} suppressing decoherence from nuclear spins  and stronger SOI than Si \cite{ge-heavy}  facilitating rapid qubit control. Ge hole qubits have shown significant advancements in recent years \cite{Ge-review,nano1,nano2,wiley,wiley2} highlighted by the demonstration of single- and two-qubit control \cite{ge-hole-exp1,ge-hole-exp2,ge-hole-exp3,ge-hole-exp4}, singlet-triplet encoding \cite{singlet-triplet}, four-qubit processor \cite{fourqubit} and successful charge control in a sixteen-dot array \cite{16qubit}. These qubits are hosted in quantum dots based on planar Ge/SiGe heterostructures, nanowires and hut wires. 

In planar Ge/SiGe quantum wells, the dot is formed by a strong confinement along the growth direction (say $z$) and weak lateral confinement created by the smoothly varying gate voltages. The low energy quasiparticles in these dots are the heavy hole states carrying effective spin $J=3/2$ \cite{Luttinger}. These states are primarily influenced by $p$-cubic Rashba SOI \cite{winkler,winkler2}, which includes cubic and spherically-symmetric terms, with the latter being more dominant. These terms arise from heavy hole (HH)/light hole (LH) mixings derived through second-order perturbation theory applied to the Luttinger-Kohn Hamiltonian \cite{Luttinger} and depend on valence band anisotropies \cite{optimal-Ge} and lateral confinement anharmonicities \cite{anharmony}.

  Recent studies also suggest the presence of $p$-linear SOIs, both Rashba and Dresselhaus types, in Ge/SiGe heterostructures \cite{emergent,Dresselhaus,special-rashba}. The $p$-linear Rashba SOI, attributed to the local $C_{2v}$ interface \cite{PhysRevB.54.5852,PhysRevB.92.165301,PhysRevB.69.115333,PhysRevB.89.075430} and determined through atomistic pseudopotential method calculations \cite{emergent}, is believed to drive electric dipole spin resonance (EDSR) in planar Ge quantum dots observed in experiments \cite{ge-hole-exp3,fourqubit} with in-plane magnetic fields. This SOI is a first-order direct Rashba effect, caused by  a combination of HH-LH mixing and a direct dipolar coupling to the external electric field \cite{emergence}. For an out-of-plane magnetic field, the less significant cubic symmetric component of $p$-cubic Rashba SOI is shown to be responsible for EDSR, resulting in slower spin rotations \cite{strained-Ge,optimal-Ge}. 
Another form of weak $p$-linear Rashba SOI has been identified \cite{Dresselhaus}, resulting from the interaction between the HH/LH manifold and remote conduction bands due to the structural inversion asymmetry of the heterostructure. A comprehensive theory of the EDSR mechanism of Ge qubits under an in-plane magnetic field has also been presented recently \cite{in-plane}. The Dresselhaus SOI was known to be absent in Ge due to its centrosymmetric structure. It has been  reported that symmetry breaking at the Ge/GeSi interfaces gives rise to a $p$-linear Dresselhaus-type  SOI \cite{Dresselhaus}, which can be stronger than cubic Rashba SOI and may dominate the behavior of quasicircular dots under out-of-plane magnetic fields, assuming the strains are uniform. Furthermore, moving the dot across inhomogeneous strain fields combined with $g$-factor modulations can induce a specific kind of $p$-linear Rashba SOI that can fasten the Rabi oscillations \cite{special-rashba}. Inhomogenous and inseparable electric fields can also induce an SOI that causes Rabi rotations under in-plane magnetic fields \cite{inhomogeneous}. 

In Ge/Si (core/shell) nanowires, the hole states have a stronger $p$-linear Direct Rashba spin-orbit interaction (DRSOI) \cite{DRSOI,circuitqed,Sercel-nanowire-1990,Csontos-2009,drsoi-growth,gfactor-wire} which can be used to leverage spin rotations about $100$ times faster than the hole qubits in planar quantum dots. Unlike the conventional Rashba coupling which arises due to structural inversion asymmetry, the DRSOI results from the dipolar coupling between the quasidegenerate ground and excited states of the nanowire under a hard-wall boundary condition along the radial direction \cite{DRSOI}. Its effect has been simulated in a  squeezed (elongated) planar Ge quantum dot and large Rabi frequencies have been reported even at small driving amplitudes \cite{squeezed}. Hence, the DRSOI holds the prospect of designing lower power ultrafast quantum gates in squeezed geometries.

The mechanism of hole spin EDSR has been theoretically investigated in both single \cite{edsr-loss-hole} and double \cite{sherman,dqd} planar Ge quantum dots. A recent study has also examined the combined effects of $p$-linear and cubic Rashba SOIs, as well as the behaviour of photoinduced Rabi oscillations under strong circular driving (beyond second-order perturbation theory) in an isotropic planar Ge dot \cite{dey}. Although the DRSOI-induced EDSR has been studied recently in squeezed dots \cite{squeezed}, the specific impact of squeezing or anisotropy in planar Ge quantum dots and its interplay with the direction of the applied electric field on the Rabi frequency has not yet been addressed yet. In this study, we examine the Rabi oscillations of an anisotropic planar Ge quantum dot under the influence of a coherent laser beam with generic polarization, considering the recently discovered $p$-linear Rashba and Dresselhaus SOIs \cite{Dresselhaus} but not the DRSOI. Although squeezing the dot may affect the SOI strengths and $g$-factors, we assume that they remain constant for the sake of simplicity \cite{hole-review}. Instead of gate voltages, we consider the driving force provided by the electric field of a coherent laser beam, as its polarization offers tunability and a broader understanding of the directional dependence of the Rabi frequency on the driving field. 

We employ both analytical and numerical approaches to study the qubit dynamics. For an out-of-plane magnetic field, we use the exact Fock-Darwin states of an elliptical potential and study the dynamics analytically using a Schrieffer-Wolff projection to the lowest Zeeman-split block. Numerical simulations using Floquet theory reveal approximate `anisotropy cutoffs,' beyond which Rabi oscillations become heavily distorted as the excited states approach the qubit block. We demonstrate that increasing anisotropy (while keeping other system parameters constant) results in a significant rise in the Rabi frequency magnitude. The Rabi frequency is enhanced when the major axes of both the ellipses align in the same direction. We also calculate Rabi frequencies for in-plane magnetic fields, commonly used in experiments, and study their variation with the rotation of the magnetic field vector on the qubit plane. We analyze the results for both Rashba and Dresselhaus SOIs, identifying the role of squeezing in determining the Rabi frequency. We derive an analytical expression showing the condition that the eccentricities of the polarization and equipotential ellipses must satisfy to achieve maximum Rabi frequency.

The paper is organized as follows. In Sec. \ref{out-B}, we discuss the physics in presence of an out-of-plane magnetic field. In Sec. \ref{Fock}, we present the theoretical model of the elliptical quantum dot and map it to the Fock-Darwin model whose eigenstates constitute the set of basis states for our problem. In Sec. \ref{Rabi-out-elec} and \ref{Rabi-out-hole}, we derive the the approximate analytical expressions of the Rabi frequency for electron- and hole-like SOIs respectively. In Sec. \ref{inplane-B}, we discuss the physics for an in-plane magnetic field. In Sec. \ref{aniso}, we model the quantum dot as an anisotropic harmonic oscillator. In Sec. \ref{Rabi-in-elec} and \ref{Rabi-in-hole}, we deduce the the approximate analytical expressions of the Rabi frequency for electron- and hole-like SOIs respectively.  In Sec. \ref{results}, we present and analyse the results of Rabi frequency for realistic system parameters and driving strengths. In Sec.s \ref{results-anal-out} and \ref{results-anal-in}, we analyse the behaviour of Rabi frequency using the analytical results obtained in Sec.s \ref{out-B} and \ref{inplane-B} for an out-of- and in-plane magnetic field respectively. In Sec. \ref{results-num}, we show results of the Rabi oscillations for the squeezing parameters where the analytical expressions of Rabi frequency are inaccurate or cannot be obtained. Finally, we conclude our results in Sec. \ref{conc}.

\begin{figure}[htbp]
		\centering
  \vspace{0.2cm}
\includegraphics[trim={0cm 3cm 0cm 5cm},clip,width=8cm]{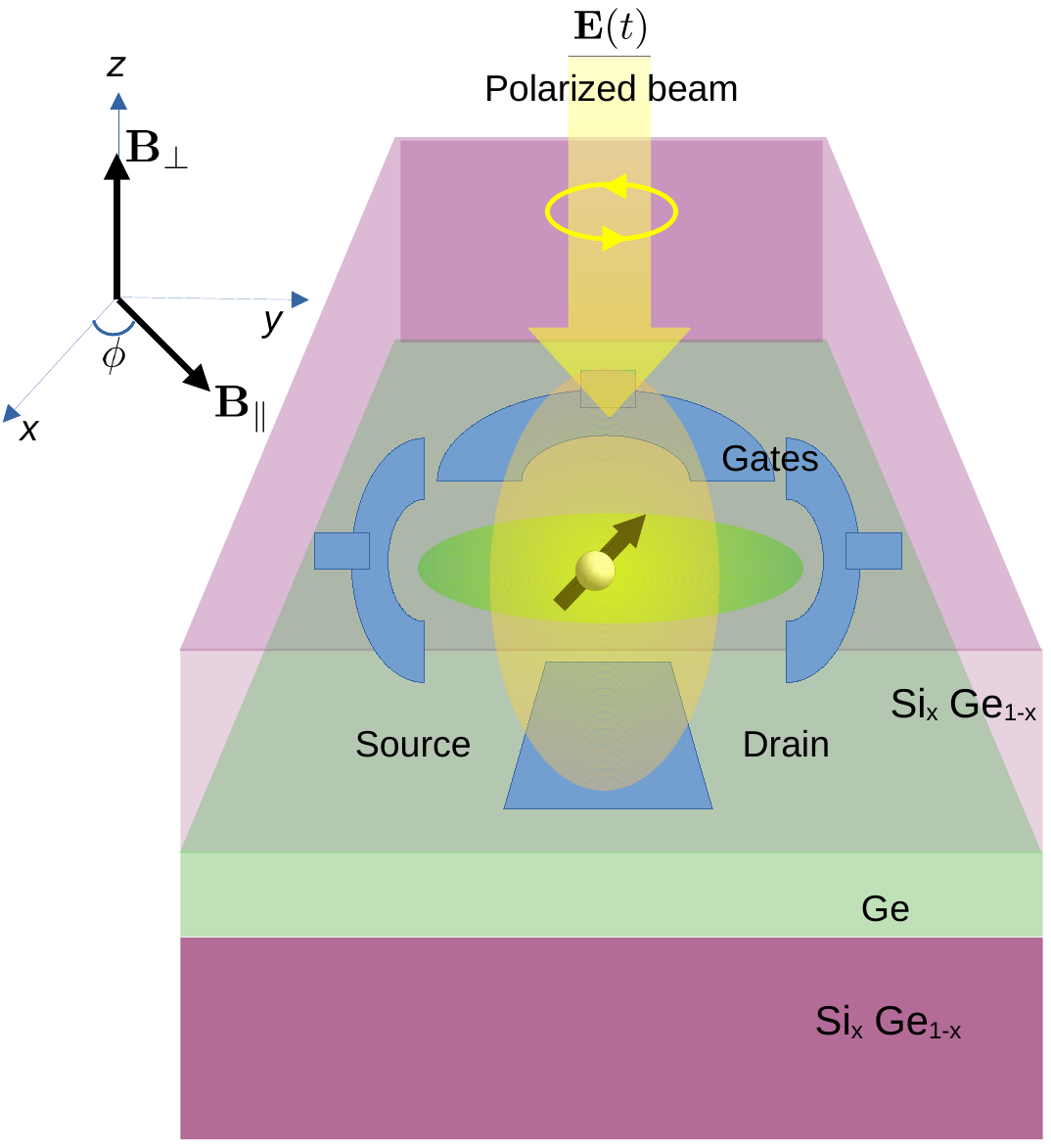}
		\caption{Schematic model of the planar Ge quantum dot where the squeezed confining potential is created by the gate electrodes and the driving pulse is applied through a polarized beam.}
		\label{schematic-model}
	\end{figure}

\section{Out-of-plane magnetic field}\label{out-B}
 \subsection{Fock-Darwin Model}\label{Fock}
 The Hamiltonian of a Ge heavy hole in an anisotropic planar quantum dot, as shown in Fig. \ref{schematic-model}, can be modelled as $H=H_0+H_\text{SOI}$ where 
 \begin{equation}\label{H0}
     H_0=\frac{p^2}{2m}+ \frac{1}{2}m(\omega_x^2 x^2+\omega_y^2 y^2)
 \end{equation}
  with $m$ being the effective heavy-hole mass and $H_{\text{SOI}}$ is the spin-orbit interaction term for the heavy-holes. The expression of $H_{\text{SOI}}$ depends on the specific type of SOI considered in the problem (which we discuss in the subsequent sections). The dimensions of the dot can be controlled using the gate voltages. Let $X_0$ and $Y_0$ be the lengths of semimajor/minor axes of the elliptical dot along the $x$ and $y$ directions, respectively. Then, the confinement frequencies are related to these dimensions as $\omega_x=\hbar/(m X_0^2)$ and $\omega_y=\hbar/(m Y_0^2)$.
  
  In presence of an out-of-plane magnetic field  ${\bf B}_\perp=(0,0,B)$, the orbital motion interacts with the field through minimal coupling ${\bf p}\rightarrow{\bf P}={\bf p}-|e|{\bf A}({\bf r})$ where ${\bf A}{\bf (r)}=B \l(-y,x\r)/2$ in the symmetric gauge and the spins couple directly with the field through the Zeeman interaction. The resulting Hamiltonian is $H_\perp=H_\text{FD}+H_{\text{Z},\perp}+H_{\text{SOI},\perp}$ where  $H_\text{FD}$ is the Fock-Darwin (FD) Hamiltonian responsible for confinement, $H_{\text{Z}}$ is the Zeeman Hamiltonian required to create the two-level spin-qubit system and $H_{\text{SOI},\perp}$ is the $B$-dependent (through minimal coupling) SOI that can cause EDSR upon periodic driving.  

 The FD Hamiltonian can be written as
 \begin{equation}\label{FD}
     H_\text{FD}=\frac{1}{2m}(p_x^2+p_y^2+\Omega_x^2 x^2+ \Omega_y^2 y^2 - m\omega_c L_z)
 \end{equation}
where $\Omega_{x,y}^2=m^2(\omega_{x,y}^2+\omega_c^2/4)$, $\omega_c=|e| B /m$ and $L_z=x p_y-y p_x$.

The above Hamiltonian is exactly solvable with the following coordinate transformations \cite{chakraborty}:
\begin{equation}
x= \cos \chi~ q_1 - \chi_2 \sin\chi~p_2,
\end{equation}
\begin{equation}
y= \cos \chi~ q_2 - \chi_2 \sin \chi~p_1  , 
\end{equation}
\begin{equation}
p_x=  \chi_1 \sin \chi~ q_2 + \cos \chi~ p_1,
\end{equation}
\begin{equation}
p_y=  \chi_1 \sin \chi~ q_1 + \cos \chi~ p_2,
\end{equation}
where $\chi_1=-\Omega/2$, $\chi_2=1/\chi_1$ and $\chi=\tan^{-1}[\sqrt{2} m \omega_c \Omega/(\Omega_x^2-\Omega_y^2)]/2$ with $\Omega=\sqrt{\Omega_x^2+\Omega_y^2}$ and $[q_i,q_j]=[p_i,p_j]=0$, $[q_i,p_j]=i\hbar \delta_{i,j}$. Upon transformation, the Hamiltonian can be simplified as
\begin{equation}\label{trans}
\begin{aligned}
    H_\text{FD}=\frac{p_1^2}{2m_1}+\frac{p_2^2}{2m_2}+\frac{1}{2}m_1\omega_1^2 q_1^2+\frac{1}{2}m_2\omega_2^2 q_2^2
\end{aligned}
\end{equation}
where $m_{1,2}=m/\alpha_{1,2}^2$ and $\omega_{1,2}=\alpha_{1,2} \beta_{1,2}/m$ with 
\begin{equation}
    \alpha_1^2=\frac{\Omega_x^2+3\Omega_y^2+\text{sgn}[\Omega_x^2-\Omega_y^2]~\Omega_3^2}{2 \Omega^2},
\end{equation}
\begin{equation}
    \alpha_2^2=\frac{3\Omega_x^2+\Omega_y^2- \text{sgn}[\Omega_x^2-\Omega_y^2] ~\Omega_3^2}{2 \Omega^2},
\end{equation}
\begin{equation}
    \beta_1^2=\frac{1}{4}\l(3\Omega_x^2+\Omega_y^2+ \text{sgn}[\Omega_x^2-\Omega_y^2]~\Omega_3^2\r),
\end{equation}
\begin{equation}
    \beta_2^2=\frac{1}{4}\l(\Omega_x^2+3\Omega_y^2-\text{sgn}[\Omega_x^2-\Omega_y^2]~\Omega_3^2\r)
\end{equation}
and
\begin{equation}
    \Omega_3^2=\l[ (\Omega_x^2 - \Omega_y^2)^2 + 2m^2\omega_c^2 \Omega^2\r]^{1/2}.
\end{equation}
Here, sgn is the signum function defined as $\text{sgn}[x]=\pm1$ for $x\gtrless0$. 

In terms of ladder operators
\begin{equation}
    a_{i}=\frac{1}{\sqrt{2}}\l(\frac{q_{i}}{\mathcal{X}_{i}}+i\frac{p_{i}}{\mathcal{P}_{i}}\r),~~ a^\dagger_{i}=\frac{1}{\sqrt{2}}\l(\frac{q_{i}}{\mathcal{X}_{i}}-i\frac{p_{i}}{\mathcal{P}_{i}}\r)
\end{equation}
with $\mathcal{X}_i=\sqrt{\hbar /(m_i \omega_i)}$, $\mathcal{P}_i=\sqrt{\hbar m_i \omega_i}$ \{$i=1,2$\}, the FD Hamiltonian can be rewritten as
\begin{equation}
    H_\text{FD}=\hbar\omega_1\left( a_1^\dagger a_1+\frac{1}{2}\right)+\hbar\omega_2\left(a_2^\dagger a_2+\frac{1}{2}\right).
\end{equation}
The Zeeman Hamiltonian can be defined as
 \begin{equation}
		\begin{aligned}
			H_{\text{Z},\perp}=-\frac{\hbar \omega_\text{Z}}{2}\sigma_z
		\end{aligned}
	\end{equation}
where $\hbar \omega_{\text{Z}}\equiv g_\perp \mu_B B$ is the Zeeman splitting with $g_\perp$ being the out-of-plane $g$-factor for the holes.

The eigenstates and eigenenergies of $H_\text{FD} + H_{\text{Z},\perp}$ are $|n_1, n_2,s\rangle$ and $E_{n_1,n_2,s}=\hbar\omega_1(n_1+\frac{1}{2})+\hbar\omega_2(n_2+\frac{1}{2})-\text{sgn}[s]\frac{\hbar \omega_\text{Z}}{2}$ respectively 
where $s=\pm3/2$ and $(n_1,n_2)$ represent the quantum numbers of the two independent harmonic oscillators along directions $(q_1,q_2)$. We shall use the FD basis $ \{|n_1, n_2, s\rangle\} $  to obtain the approximate analytical and exact numerical solutions to the time dependent Schr\"{o}dinger equation upon driving by a coherent laser in presence of out-of-plane ${\bf B}$. 

\subsection{EDSR with electron-like Rashba SOI}\label{Rabi-out-elec}

The conventional SOI known for the heavy holes in planar Ge heterostructures is $p$-cubic Rashba, given by 
\begin{equation}
    H_\text{SOI}^c=i\alpha_R^{(1)} p_+ p_- p_+ \sigma_+ + i\alpha_R^{(2)}p_+^3 \sigma_- +\text{H.c.},
\end{equation}
 where $\alpha_R^{(1)}$ and $\alpha_R^{(2)}$ are the coefficients of the cubic- and spherically-symmetric terms respectively \cite{winkler,winkler2}. The Dresselhaus SOI is absent due to bulk inversion symmetry of Ge. For an out-of-plane magnetic field, EDSR can occur only if $\alpha_R^{(1)}\neq0$ \cite{strained-Ge,optimal-Ge}. However, the magnitude of $\alpha_R^{(1)}$ is very small in these systems which leads to extremely low Rabi frequencies. Hence, we ignore the $p$-cubic Rashba coupling for the rest of the paper. In Ref. \cite{emergent}, it has been reported that the SOI responsible for the EDSR observed in experiments with planar Ge quantum dots is of $p$-linear Rashba type, which has the form
\begin{equation}\label{elike}
    H_\text{SOI}^l=-i\alpha_l(p_-\sigma_+-p_+\sigma_-).
\end{equation}
This Rashba SOI has similar form as that of the conduction electrons and hence we term it as `electron-like' Rashba SOI.
 For an out-of-plane magnetic field, the SOI also becomes $B$-dependent through minimal coupling and simplifies as 
\begin{equation}
\begin{aligned}
   H_{\text{SOI},\perp}^l=&\alpha_l\l(-f^{(+)}_{1-} a_1 + f^{(-)}_{1-} a_1^\dagger+i f_{2+}^{(-)} a_2 -i f_{2+}^{(+)} a_2^\dagger\r)\sigma_+ \\
   & + \text{H.c.}
\end{aligned}
\end{equation}
Here, $f^{(a)}_{bc}$ are real-valued functions defined as $f^{(\pm)}_{i\pm}=f^{(\mathcal{P})}_{i\pm}\pm f^{(\mathcal{X})}_{i\pm}$ with
\begin{equation}\label{f_1}
    f^{(\mathcal{P})}_{i\pm}=\frac{\mathcal{P}_i}{\sqrt{2}}\l(\cos \chi \pm \frac{m \omega_c \chi_2 \sin \chi}{2}\r)
\end{equation}
and 
\begin{equation}\label{f_2}
    f^{(\mathcal{X})}_{i\pm}=\frac{\mathcal{X}_i}{\sqrt{2}}\l(\chi_1 \sin \chi \pm \frac{m \omega_c \cos\chi}{2}\r).
\end{equation}

Hence, the total Hamiltonian of the heavy hole in presence of out-of-plane magnetic field and electron-like Rashba SOI can be written as $H^l_\perp=H_\text{FD}+H_{\text{Z},\perp}+H_{\text{SOI},\perp}^l$. To observe EDSR, we drive the system with an electrical pulse provided by a coherent laser beam. 

Let us consider a beam of generic polarization  incident normally on the planar dot with the electric field vector ${\bf E}({\bf r},t)=[E_{0x}\sin(\omega t + kz),E_{0y}\cos (\omega t + kz),0]$. Then, the driving potential at the quantum dot plane ($z=0$) can be written in the length gauge as \cite{dey}
\begin{equation}
    V({\bf r},t)=-|e|\int_{\bf r} {\bf E}\cdot d{\bf r}^\prime=-(F_{0x} x \sin \omega t + F_{0y} y \cos \omega t)
\end{equation}
where $F_{0 x,y}=|e|E_{0 x,y}$. In terms of ladder operators, we have
\begin{equation}
    V({\bf r},t)=v_1(t) a_1+ v_2(t) a_2 + \text{H.c.},
\end{equation}
where
\begin{equation}
    v_1(t)=-\frac{1}{\sqrt{2}}\l(\mathcal{X}_1 F_{0x}\sin \omega t \cos \chi+i \mathcal{P}_1 F_{0y}\cos \omega t ~\chi_2 \sin \chi\r)
\end{equation}
and 
\begin{equation}
    v_2(t)=-\frac{1}{\sqrt{2}}\l(i \mathcal{P}_2 F_{0x}\sin \omega t ~\chi_2 \sin \chi + \mathcal{X}_2 F_{0y}\cos \omega t \cos \chi\r).
\end{equation}

The total Hamiltonian including the periodic drive is $H^l_\perp+V({\bf r},t)$. On performing a Schrieffer-Wolff transformation [see Appendix \ref{SWT-app}], we get an effective EDSR Hamiltonian for the qubit as 
\begin{equation}\label{Heffl}
\begin{aligned}
    [H^l_{\perp}]_\text{eff}(t)&=-\l(\frac{\hbar \omega_\text{Z}+\Delta^l_\perp}{2}\r)\sigma_z\\
    &+\frac{\hbar }{2} (\omega_{\text{res},\perp}^l \e^{i \omega t} +\omega_{\text{off},\perp}^l\e^{-i \omega t}) \sigma_+ + \text{H.c.}
    \end{aligned}
\end{equation}
where
\begin{equation}\label{rabiformula}
\begin{aligned}
&\omega_{\text{res},\perp}^l=\frac{1}{\sqrt{2}\chi_1} \times\\
&\bigg\{\chi_1 \cos \chi\l[i \mathcal{X}_1F_{0x} (S^{(2)}_{1a}-S^{(2)}_{1b}) +\mathcal{X}_2F_{0y} (-S^{(2)}_{2a}+S^{(2)}_{2b})\r]\\
&+\sin\chi\l[ i \mathcal{P}_1 F_{0y}  (S^{(2)}_{1a}+S^{(2)}_{1b})+ \mathcal{P}_2F_{0x} (S^{(2)}_{2a}+S^{(2)}_{2b}) \r]\bigg\},
\end{aligned}
\end{equation}
\begin{equation}\label{off}
\begin{aligned}
&\omega_{\text{off},\perp}^l=\frac{1}{\sqrt{2}\chi_1} \times\\
&\bigg\{\chi_1 \cos \chi\l[-i \mathcal{X}_1F_{0x} (S^{(2)}_{1a}-S^{(2)}_{1b}) +\mathcal{X}_2F_{0y} (-S^{(2)}_{2a}+S^{(2)}_{2b})\r]\\
&+\sin\chi\l[ i \mathcal{P}_1 F_{0y}  (S^{(2)}_{1a}+S^{(2)}_{1b})- \mathcal{P}_2F_{0x} (S^{(2)}_{2a}+S^{(2)}_{2b}) \r]\bigg\}
\end{aligned}
\end{equation}
and
\begin{equation}\label{Delperp}
\begin{aligned}
    \Delta_\perp^l=\frac{\alpha_l^2}{\hbar}\l[\frac{(f_{1-}^{(+)})^2}{\omega_1+\omega_\text{Z}} + \frac{(f_{2+}^{(-)})^2}{\omega_2+\omega_\text{Z}}-\frac{(f_{1-}^{(-)})^2}{\omega_1-\omega_\text{Z}} - \frac{(f_{2+}^{(+)})^2}{\omega_2-\omega_\text{Z}}\r].
\end{aligned}
\end{equation}
The expressions of $S^{(2)}_{1a}, S^{(2)}_{1b}, S^{(2)}_{2a}$ and $S^{(2)}_{2b}$ are provided in Appendix \ref{SWT-app}. For $\omega_{\text{off}}^\perp\ll  \omega_\text{Z}$, the term $\propto\e^{i\omega t}$ in Eq. (\ref{Heffl}) contributes to the Rabi oscillations with resonant frequency $|\omega^l_{\text{res},\perp}|$  while the term $\propto\e^{-i\omega t}$ gives the rapidly oscillating contributions which can be discarded by the rotating wave approximation. The resonance condition is $\omega=\omega_\text{Z}+\Delta_\perp^l/\hbar$. 

The orientation of the ellipse of polarization can also be varied on the $x$-$y$ plane (keeping the centre fixed). Let the ellipse be rotated through some angle $\theta$ about the $z$-axis of the squeezed confinement. We label $\theta$ as the `orientation' angle. The electric field then transforms as ${\bf E}_\theta ({\bf r},t)= R_\theta {\bf E} ({\bf r}, t)$  where $R_\theta$ is the standard rotation matrix about the $z$-axis defined as
\begin{equation}
    R_\theta=\l(\begin{array}{cc}
      \cos \theta   & -\sin \theta \\
        \sin \theta & \cos \theta
    \end{array}\r).
\end{equation}
Then, the resonant Rabi frequency for an orientation angle $\theta$ is given by $|\omega_{\text{res},\perp}^l(\theta)|$ where
\begin{equation}
\begin{aligned}
    &\omega_{\text{res},\perp}^l(\theta)=\omega_{\text{res},\perp}^l \cos \theta +\frac{\sin \theta}{\sqrt{2}\chi_1} \times\\
&\bigg\{\chi_1 \cos \chi\l[\mathcal{X}_1F_{0y} (S^{(2)}_{1a}-S^{(2)}_{1b}) +i\mathcal{X}_2F_{0x} (S^{(2)}_{2a}-S^{(2)}_{2b})\r]\\
&+\sin\chi\l[  \mathcal{P}_1 F_{0x}  (S^{(2)}_{1a}+S^{(2)}_{1b})- i\mathcal{P}_2F_{0y} (S^{(2)}_{2a}+S^{(2)}_{2b}) \r]\bigg\}
\end{aligned} 
\end{equation}
where $\omega_{\text{res},\perp}^l$ is defined in equation (\ref{rabiformula}).

\subsection{EDSR with hole-like Dresselhaus and Rashba SOI}\label{Rabi-out-hole}

In Ref. \cite{Dresselhaus}, $p$-linear Dresselhaus $(H_\text{D}^{(+)})$ and Rashba $(H_\text{R}^{(+)})$ SOIs have been derived for heavy holes in planar Ge/Si heterostructures where
\begin{equation}
    H_\text{D}^{(+)}=\alpha_D(p_x\sigma_x+p_y\sigma_y)=\alpha_D(p_-\sigma_++p_+\sigma_-)
\end{equation}
and
\begin{equation}\label{Rashbasoc2}
H_\text{R}^{(+)}=\alpha_R(p_x\sigma_y+p_y\sigma_x)=-i\alpha_R(p_+\sigma_+-p_-\sigma_-)
\end{equation}
such that the net SOI is 
\begin{equation}\label{hlike}
\begin{aligned}
    H_{\text{SOI}}^{(+)}&=\alpha_D(p_-\sigma_+ + p_+\sigma_-) -i \alpha_R (p_+\sigma_+-p_-\sigma_-)\\
 &=(\alpha_D p_- -i \alpha_R p_+)\sigma_+ + \text{H.c.}
\end{aligned}
\end{equation}
Here, the `$+$' sign replaces the conventional `$-$' sign between the $\sigma_x$ and $\sigma_y$ terms present for electrons in the Rashba or Dresselhaus SOIs  because spin $3/2$ transforms differently from spin $1/2$ under certain symmetry operations \cite{Dresselhaus}.

In presence of an out-of-plane magnetic field, $p_\pm \rightarrow P_\pm=P_x \pm i P_y$ and hence we get the $B$-dependent hole-like SOI as
\begin{equation}
     H_{\text{SOI},\perp}^{(+)}=(h_{1a} a_1 + h_{1b} a^\dagger_1 + h_{2a} a_2 + h_{2b} a^\dagger_2)\sigma_+ + \text{H.c.}
\end{equation}
 where
 \begin{equation}
     h_{1a}=-\l(i\alpha_D f_{1-}^{(+)}+\alpha_R f_{1-}^{(-)}\r)
 \end{equation}
 \begin{equation}
     h_{1b}=i\alpha_D f_{1-}^{(-)}+\alpha_R f_{1-}^{(+)}
 \end{equation}
\begin{equation}
     h_{2a}=-\l(\alpha_D f_{2+}^{(-)}+i\alpha_R f_{2+}^{(+)}\r)
 \end{equation}
\begin{equation}
     h_{2b}=\alpha_D f_{2+}^{(+)}+i\alpha_R f_{2+}^{(-)}
 \end{equation}
 where $f^{(a)}_{bc}$ are defined in Eqs. (\ref{f_1}) and (\ref{f_2}). Using a Schrieffer-Wolff transformation and driving with $V({\bf r}, t)$, we get the effective EDSR Hamiltonian as
\begin{equation}
\begin{aligned}
    [H_{\perp}^{(+)}]_\text{eff}(t)=&-\l(\frac{\hbar \omega_\text{Z} + \Delta_\perp^{(+)}}{2}\r)\sigma_z\\ &+\l[\frac{\hbar}{2}(\omega_{\text{res},\perp}^{(+)} \e^{i\omega t}+\omega_{\text{off},\perp}^{(+)} \e^{-i\omega t})\sigma_+ + \text{H.c.}\r]
\end{aligned}
\end{equation}
where
\begin{equation}
\begin{aligned}
    \Delta_\perp^{(+)}&= \frac{1}{\hbar }\l[\frac{|h_{1a}|^2}{\omega_1+\omega_\text{Z}} +  \frac{|h_{2a}|^2}{ \omega_2+\omega_\text{Z}}-\frac{|h_{1b}|^2}{\omega_1- \omega_\text{Z}} -  \frac{|h_{2b}|^2}{ \omega_2- \omega_\text{Z}}\r],
\end{aligned}
\end{equation}
 and $\omega_{\text{res},\perp}^{(+)}$ and $\omega_{\text{off},\perp}^{(+)}$  have same expressions as  $\omega^l_{\text{res},\perp}$ and $\omega^l_{\text{off},\perp}$ in (\ref{rabiformula}) and (\ref{off}) respectively but with new $\{S^{(2)}_{lm}\}$ defined as:
 \begin{equation}
     S^{(2)}_{1a}=-\frac{h_{1a}}{\hbar \omega_1+\hbar \omega_\text{Z}},
 \end{equation}
  \begin{equation}
     S^{(2)}_{1b}=\frac{h_{1b}}{\hbar \omega_1-\hbar \omega_\text{Z}},
 \end{equation}
  \begin{equation}
     S^{(2)}_{2a}=-\frac{h_{2a}}{\hbar \omega_2+\hbar \omega_\text{Z}}
 \end{equation}
 and
  \begin{equation}
     S^{(2)}_{2b}=\frac{h_{2b}}{\hbar \omega_2-\hbar \omega_\text{Z}}.
 \end{equation}
The resonant Rabi frequency is $|\omega_{\text{res},\perp}^{(+)}|$ and the resonance condition is $\omega=\omega_\text{Z}+\Delta_\perp^{(+)}/\hbar$.

\section{In-plane magnetic field}\label{inplane-B}

\subsection{Model}\label{aniso}

Let us consider a generic in-plane magnetic field which makes an angle $\phi$ with the $x$-axis i.e. ${\bf B}=(B_x,B_y,0)=B(\cos \phi, \sin \phi,0)$. The vector potential can be chosen as ${\bf A}({\bf r})=B(0,0, y\cos \phi - x \sin \phi)$, which does not couple to the orbital degree of freedom as the out of plane motion of the hole is quenched.  Then, the 2D heavy-hole Hamiltonian is $H_{||}=H_0+H_{\text{Z},||}+H_\text{SOI}$ where $H_0$ is defined in (\ref{H0}), $H_\text{SOI}$ can be electron- or hole-like as defined in Eqs. (\ref{elike}) and (\ref{hlike}) respectively, and
\begin{equation}\label{Hzphi}
   H_{\text{Z},||}=-\frac{g_{||}\mu_B}{2}(\sigma_x B_x - \sigma_y B_y)=-\frac{\hbar \omega_\text{Z}}{2}\l(\e^{i\phi} \sigma_+  + \e^{-i\phi} \sigma_-\r) 
\end{equation}
with $\omega_\text{Z}=g_{||}\mu_B B$. Thus the in-plane $g$-factor is anisotropic i.e. $g_{yy}=-g_{xx}=-g_{||}$.
Consequently, the spin vector $\langle\boldsymbol{\sigma}(t)\rangle $ of a heavy hole makes an angle $2\phi$ or $\pi-2\phi$ with the direction of ${\bf B}$ in the $|+\rangle$ or $|-\rangle$ eigenstates respectively. This is in contrast with the electronic qubits where the spin vector of $|\pm\rangle$ states are aligned along/opposite to ${\bf B}$. 

Since we want to deduce an effective 2-level Rabi Hamiltonian upon driving by the laser, we first diagonalize $(\ref{Hzphi})$ by the unitary transformation: $\tilde{H}_{\text{Z},||}=U^\dagger H_{\text{Z},||} U =  -\frac{\hbar \omega_\text{Z}}{2} \sigma_z$ where 
\begin{equation}
    U=\frac{1}{\sqrt{2}}\l ( \begin{array}{cc}
        1 & 1 \\
        \e^{-i\phi} & -\e^{-i\phi}
    \end{array} \r).
\end{equation}
Similarly, $\tilde{H}_0=U^\dagger H_0 U=H_0$. In terms of ladder operators
\begin{equation}
    a_x=\frac{1}{\sqrt{2}}\l(\frac{x}{X_0}+i\frac{p_x}{P_{x0}}\r),~~ (a_{x})^\dagger=a_x^\dagger
\end{equation}
and
\begin{equation}
    a_y=\frac{1}{\sqrt{2}}\l(\frac{y}{Y_0}+i\frac{p_y}{P_{y0}}\r),~~ (a_{y})^\dagger=a_y^\dagger
\end{equation}
with $X_0=\sqrt{\hbar /(m \omega_x)}$, $P_{x0}=\sqrt{\hbar m \omega_x}$, $Y_0=\sqrt{\hbar /(m \omega_y)}$ and $P_{y0}=\sqrt{\hbar m \omega_y}$, we can write
\begin{equation}
     \tilde{H}_0=\hbar\omega_x\l(a_x^\dagger a_x+\frac{1}{2}\r)+\hbar\omega_y\l(a_y^\dagger a_y+\frac{1}{2}\r).
\end{equation}
Hence, the eigenstates and eigenvalues of $\tilde{H}_0+\tilde{H}_{\text{Z},||}$ are $|n_x, n_y,s\rangle$ and $E_{n_x,n_y,s}=\hbar\omega_1(n_x+\frac{1}{2})+\hbar\omega_2(n_y+\frac{1}{2})-\text{sgn}[s]\frac{\hbar \omega_\text{Z}}{2}$ respectively 
where $s=\pm3/2$ and $(n_x,n_y)$ represent the quantum numbers of the two uncoupled harmonic oscillators along directions $(x,y)$. For in-plane magnetic field, we shall use the oscillator basis $ \{|n_x, n_y, s\rangle\} $ later to obtain the approximate analytical and exact numerical results of the Rabi frequency.

\subsection{EDSR with electron-like Rashba SOI}\label{Rabi-in-elec}

For the electron-like Rashba SOI of Eq. (\ref{elike}), the unitary transformation yields
\begin{equation}
\begin{aligned}
    \tilde{H}_\text{SOI}^l&=U^\dagger H_\text{SOI}^lU=\frac{-i \alpha_l}{2}\big[\l( p_- \e^{-i\phi} - p_+ \e^{i\phi}\r)\sigma_z \\
    &~~+\l( p_- \e^{-i\phi} + p_+ \e^{i\phi}\r)\l(\sigma_- - \sigma_+\r)\big]\\
    &=-\frac{\alpha_l}{\sqrt{2}}\bigg[i\l\{\mathcal{P}_x\sin \phi (a_x^\dagger-a_x) + \mathcal{P}_y\cos \phi (a_y^\dagger-a_y)\r\}\sigma_z\\
    &+\l\{\mathcal{P}_x \cos\phi (a_x^\dagger-a_x)-\mathcal{P}_y \sin\phi (a_y^\dagger-a_y)\r\}\l(\sigma_+ - \sigma_-\r)\bigg]
\end{aligned}
\end{equation}
Similarly, the drive $\tilde{V}({\bf r},t)=U^\dagger V({\bf r},t) U=V({\bf r},t)$ can be written as
\begin{equation}\label{vt}
    \tilde{V}({\bf r},t)=-\frac{1}{\sqrt{2}}\bigg[F_{0x} X_0 (a_x^\dagger + a_x) \sin \omega t + F_{0y} Y_0 (a_y^\dagger + a_y) \cos \omega t \bigg].
\end{equation}
The total Hamiltonian with driving is hence $\tilde{H}_{||}^l=\tilde{H}_0+ \tilde{H}_{Z,||}+\tilde{H}_\text{SOI}^l+\tilde{V}({\bf r},t)$.
Again, performing SW transformation, the effective EDSR Hamiltonian of the qubit is obtained as
\begin{equation}
\begin{aligned}
    [H_{||}^l]_\text{eff}=&-\l(\frac{\hbar\omega_\text{Z}+\Delta_{||}^l (\phi)}{2}\r)\sigma_z\\&+\frac{\hbar}{2}[~\omega^l_{\text{res},||} (\phi) \e^{i \omega t} 
    + \text{H.c.}~]\sigma_+ +~ \text{H.c.}
\end{aligned}
\end{equation}
where
\begin{equation}
\begin{aligned}
    \Delta_{||}^l (\phi)=\frac{\alpha_R^2 }{2\hbar}&\bigg[P_{0x}^2 \cos^2 \phi\l( \frac{1}{\omega_x+\omega_\text{Z}} - \frac{1}{\omega_x-\omega_\text{Z}}\r)\\
    &+P_{0y}^2 \sin^2 \phi\l( \frac{1}{\omega_y+\omega_\text{Z}} - \frac{1}{\omega_y-\omega_\text{Z}}\r)\bigg]
\end{aligned}
\end{equation}
and
\begin{equation}\label{rabi-inplane-gen}
\begin{aligned}
   \omega_{\text{res}, ||}^l  (\phi)=\frac{i\alpha_R }{2 \hbar }&\bigg[ F_{0x} \cos\phi\l( \frac{1}{\omega_x-\omega_\text{Z}} - \frac{1}{\omega_x+\omega_\text{Z}}\r) \\ & -i F_{0y}\sin \phi\l( \frac{1}{\omega_y-\omega_\text{Z}} - \frac{1}{\omega_y+\omega_\text{Z}}\r) \bigg]. 
\end{aligned}
\end{equation}
Thus, the resonant Rabi frequency is 
\begin{equation}\label{rabi-big}
\begin{aligned}
     & |\omega_{\text{res}, ||}^l  (\phi)|=\frac{\alpha_R \omega_\text{Z}}{\hbar}\l[\frac{F_{0x}^2 \cos^2 \phi}{(\omega_{x}^2-\omega_\text{Z}^2)^2} + \frac{F_{0y}^2 \sin^2 \phi}{(\omega_{y}^2-\omega_\text{Z}^2)^2}\r]^{1/2}
\end{aligned}
\end{equation}
For a linearly polarized radiation,  $|\omega_{\text{res}, ||}^l  (\phi)|$ vanishes if ${\bf E}(t)\perp{\bf B}$ and is maximum when ${\bf E}(t) ~ || ~{\bf B}$. For $x$-polarized beams, $|\omega_{\text{res}, ||}^l  (\phi)|$ peaks when ${\bf B}||\hat{e}_x$ and $\omega_x \gtrsim \omega_\text{Z}$ whereas for $y$-polarized beams, $|\omega_{\text{res}, ||}^l  (\phi)|$ peaks when ${\bf B}||\hat{e}_y$ and $\omega_y\gtrsim \omega_\text{Z}$. The transformation $\gamma \to \gamma\pm \pi$ change the sense of rotation of elliptical polarization, while $\phi \to \phi\pm \pi$ flips the direction of ${\bf B}$. We observe that $|\omega_{\text{res}, ||}^l (\phi)|$ is independent of the sense of rotation of ${\bf E}(t)$ and ${\bf B}$-flip operation. 

For $\omega_x,\omega_y\gg \omega_\text{Z}$ i.e. stronger confinement (smaller quantum dots) or low magnetic fields, the Rabi frequency is approximately
\begin{equation}
    |\omega_{\text{res}, ||}^l  (\phi)|\approx\frac{\alpha_R \omega_\text{Z}}{\hbar}\l[\frac{F_{0x}^2 }{\omega_{x}^4} \cos^2 \phi + \frac{F_{0y}^2 }{\omega_{y}^4}\sin^2 \phi\r]^{1/2}.
\end{equation}
In such cases, if $F_{0x}/F_{0y}={\omega_x^2}/{\omega_y^2}$, then $\omega_{\text{res}, ||}^l\approx \alpha_R F_{0x} \omega_\text{Z}/(\hbar \omega_x^2) =\alpha_R F_{0y} \omega_\text{Z}/(\hbar \omega_y^2)$ is independent of the orientation of ${\bf B}$. In other words, if the major axes of the polarization and potential ellipses are perpendicular to each other and their eccentricities $e_{E}$ and $e_{C}$ (respectively) satisfy the relation $\sqrt{1-e_E^2}=1-e_C^2$, the Rabi-frequency is $\phi$-independent.

\subsection{EDSR with hole-like Rashba and Dresselhaus SOIs}\label{Rabi-in-hole}

For the hole-like Dresselhaus and Rashba SOIs of Eq. (\ref{hlike}), the unitary transformation yields
\begin{equation}
\begin{aligned}
    &\tilde{H}_\text{SOI}^{(+)}=\frac{i}{\sqrt{2}}\sigma_z [P_{0x} (\alpha_D \cos \phi-\alpha_R \sin \phi)(a_x^\dagger-a_x)\\
    &+P_{0y} (-\alpha_D \sin \phi+\alpha_R \cos \phi) (a_y^\dagger-a_y) ]\\
    &-\frac{1}{\sqrt{2}}(\sigma_+-\sigma_-)[P_{0x}(\alpha_D \sin \phi + \alpha_R \cos \phi)(a_x^\dagger-a_x)\\
    &+P_{0y}(\alpha_D \cos \phi + \alpha_R \sin \phi)(a_y^\dagger-a_y)].
\end{aligned}
\end{equation}

The total Hamiltonian with driving is hence $\tilde{H}_{||}^{(+)}=\tilde{H}_0+ \tilde{H}_{Z,||}+\tilde{H}_\text{SOI}^{(+)}+\tilde{V}({\bf r},t)$.
Again, performing SW transformation, the effective EDSR Hamiltonian of the qubit is obtained as
\begin{equation}
\begin{aligned}
    [H_{||}^l]_\text{eff}=&-\l(\frac{\hbar\omega_\text{Z}+\Delta_{||}^{(+)} (\phi)}{2}\r)\sigma_z\\&+\frac{\hbar}{2}[~\omega^{(+)}_{\text{res},||} (\phi) ~\e^{i \omega t} 
    + \text{H.c.}~]\sigma_+ +~ \text{H.c.}
\end{aligned}
\end{equation}
where 
\begin{equation}
\begin{aligned}
   & \omega^{(+)}_{\text{res},||} (\phi)=\frac{i\omega_\text{Z}}{\hbar}\times\\
    &~~\l[\frac{F_{0x}(\alpha_D \sin \phi +\alpha_R \cos \phi)}{\omega_x^2-\omega_\text{Z}^2}+i\frac{F_{0y}(\alpha_D \cos \phi +\alpha_R \sin \phi)}{\omega_y^2-\omega_\text{Z}^2}\r]
\end{aligned}
\end{equation}
and \begin{equation}
    \begin{aligned}
        &\Delta_{||}^{(+)} (\phi) \\&=\frac{1 }{2\hbar}\bigg[P_{0x}^2 (\alpha_D \sin \phi+\alpha_R \cos \phi)^2\l( \frac{1}{\omega_x+\omega_\text{Z}} - \frac{1}{\omega_x-\omega_\text{Z}}\r)\\
    &~~~~~+P_{0y}^2 (\alpha_D \cos \phi+\alpha_R \sin \phi)^2\l( \frac{1}{\omega_y+\omega_\text{Z}} - \frac{1}{\omega_y-\omega_\text{Z}}\r)\bigg]\\
    &=-\frac{1 }{\hbar}\bigg[ \frac{P_{0x}^2 \omega_\text{Z}}{\omega_x^2-\omega_\text{Z}^2} (\alpha_D \sin \phi+\alpha_R \cos \phi)^2\\
    &~~~~~~~~~~~\frac{P_{0y}^2 \omega_\text{Z}}{\omega_y^2-\omega_\text{Z}^2} (\alpha_D \cos \phi+\alpha_R \sin \phi)^2\bigg].
    \end{aligned}
\end{equation}
The resonant Rabi frequency is
\begin{equation}\label{rabi-inplane-hlike}
\begin{aligned}
    &|\omega^{(+)}_{\text{res},||} (\phi)|=\frac{\omega_\text{Z}}{\hbar}\times\\
    &~~\l[\frac{F^2_{0x}(\alpha_D \sin \phi +\alpha_R \cos \phi)^2}{(\omega_x^2-\omega_\text{Z}^2)^2}+\frac{F^2_{0y}(\alpha_D \cos \phi +\alpha_R \sin \phi)^2}{(\omega_y^2-\omega_\text{Z}^2)^2}\r]^{1/2}
\end{aligned}  
\end{equation}
Resonance condition is $\omega=\omega_\text{Z}+\Delta_{||}^{(+)} (\phi)/\hbar$.

We find that the expression, and hence the behaviour, of the resonant Rabi frequency is identical for purely electron- and hole-like $p$-linear Rashba SOIs (i.e. $\alpha_D=0$). For a purely hole-like Dresselhaus SOI (i.e. $\alpha_R=0$), on irradiation by a linearly polarized beam, the Rabi frequency vanishes if ${\bf E}(t)~||~{\bf B}$ and is maximum if ${\bf E}(t) \perp {\bf B}$. For $x$-polarized beams, Rabi frequency peaks when ${\bf B}||\hat{e}_y$ and $\omega_x \gtrsim \omega_\text{Z}$ whereas for $y$-polarized beams, it peaks when ${\bf B}||\hat{e}_x$ and $\omega_y \gtrsim \omega_\text{Z}$. Similar to the case of Rashba SOI, the Rabi frequency does not change on flipping ${\bf B}$ or the sense of rotation of ${\bf E}(t)$. Thus, the behaviour of the Rabi frequency for hole-like Dresselhaus SOI has stark differences from that of Rashba SOI. These features can hence act as probes to detect the nature of $p$-linear SOI present in the planar heterostructure and also estimate their relative strengths.

For circularly polarized radiation $(F_{0x}=F_{0y}=F_0)$ and isotropic confinement $(\omega_x=\omega_y=\omega_0)$, we deduce the Rabi frequency from equation (\ref{rabi-inplane-hlike}) as
\begin{equation}\label{rabi-inplane-hlike-iso}
\begin{aligned}
    \l[|\omega^{(+)}_{\text{res},||} (\phi)|\r]_\text{cir,iso}&=\frac{\omega_\text{Z} F_0}{\hbar(\omega_0^2-\omega_\text{Z}^2)}\times\\
    &~~\l[\alpha_D^2+\alpha_R^2+2\alpha_R \alpha_D \sin 2\phi \r]^{1/2}
\end{aligned}  
\end{equation}
The above equation shows that the Rabi frequency is $\pi$-periodic in $\phi$ with the maximum value $\omega_\text{Z} F_0 (\alpha_D+\alpha_R)/\hbar(\omega_0^2-\omega_\text{Z}^2)$ at $\phi=\pi/4$ and minimum value $\omega_\text{Z} F_0 |\alpha_D-\alpha_R|/\hbar(\omega_0^2-\omega_\text{Z}^2)$ at $\phi=3\pi/4$. A similar $\phi$ dependence can be seen for a general polarization and confinement when $\alpha_D=\alpha_R=\alpha$,
\begin{equation}\label{rabi-inplane-hlike-alpha}
\begin{aligned}
    &\l[|\omega^{(+)}_{\text{res},||} (\phi)|\r]_\alpha=\frac{\omega_\text{Z}\alpha}{\hbar}\times\\
    &~~\l[\l(\frac{F^2_{0x}}{(\omega_x^2-\omega_\text{Z}^2)^2}+\frac{F^2_{0y}}{(\omega_y^2-\omega_\text{Z}^2)^2}\r)\l(1+\sin 2\phi\r)\r]^{1/2}.
\end{aligned}  
\end{equation}
In this case, no Rabi oscillations occur when $\phi=3\pi/4$.

\begin{figure}[htbp]
		\centering
  \vspace{0.2cm}
\includegraphics[trim={0cm 0cm 0cm 0cm},clip,width=7cm]{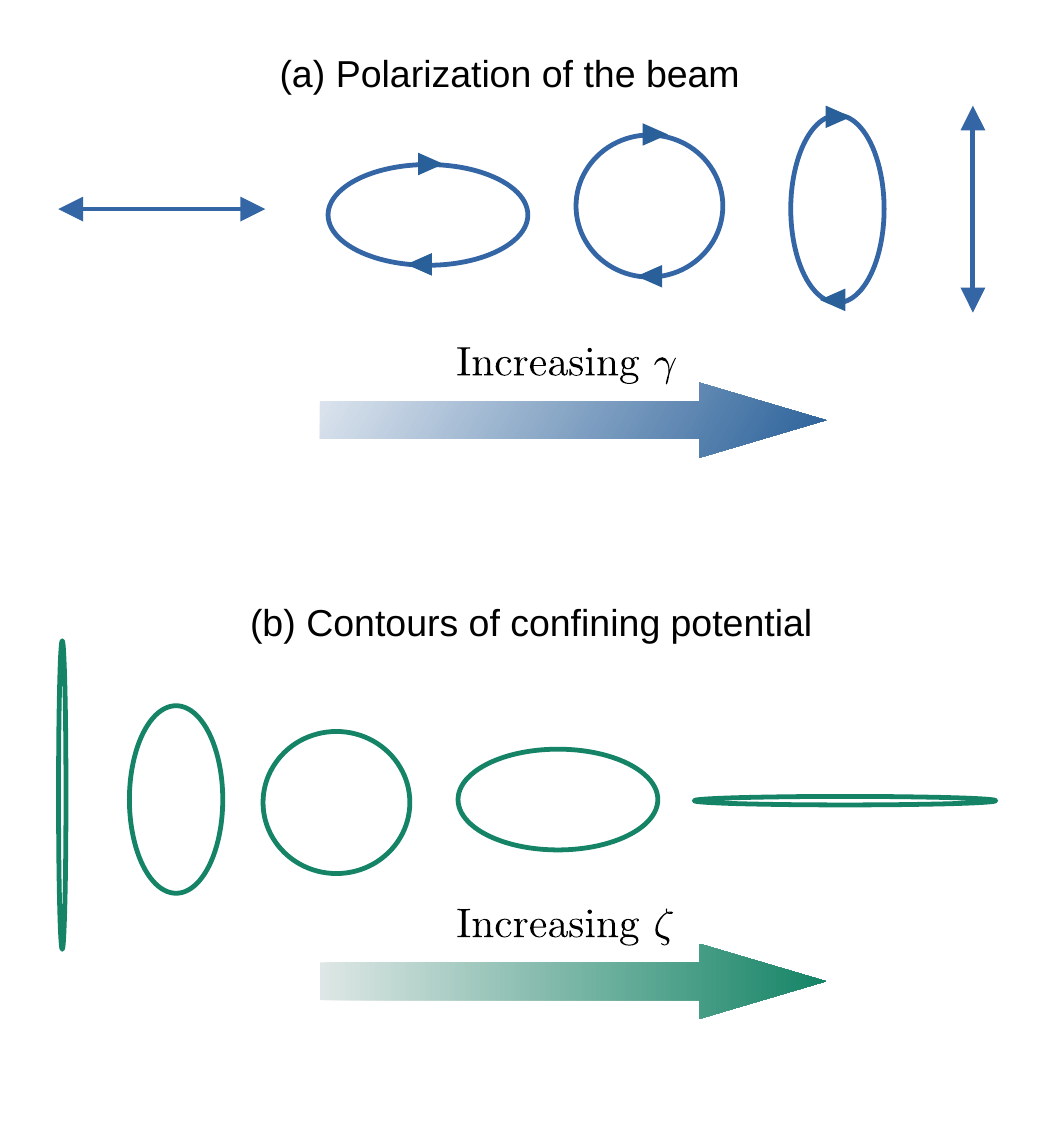}
		\caption{Schematic representation of the variation of polarization of the beam and contours of the confining potential with $\gamma$ and $\zeta$ respectively.}
		\label{schematic}
	\end{figure}

\begin{figure}[htbp]
		\centering
		\hspace{-0.3cm}\includegraphics[trim={1cm 0cm 0cm 0cm},clip,width=9cm]{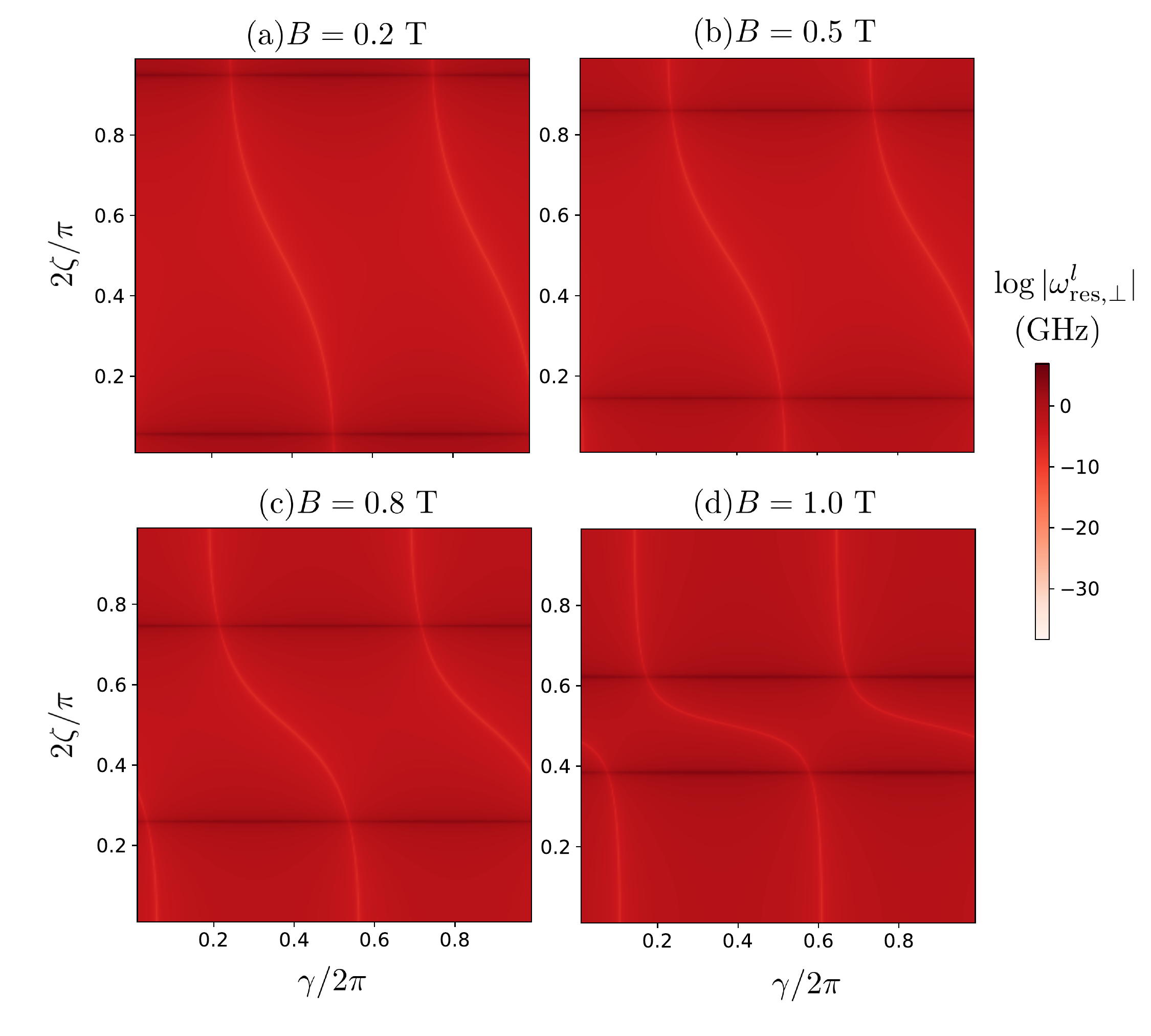}
		\caption{Density plot of the natural logarithm of the resonant Rabi frequency as a function of $\gamma$ (polarization angle) and $\zeta$ (squeezing angle) for (a)$B=0.2$ T, (b) $B=0.5$ T, (c)$B=0.8$ T and (d) $B=1$ T. }
		\label{gamma-zeta}
	\end{figure}
\begin{figure}[htbp]
		\centering
  \vspace{0.2cm}
\includegraphics[trim={1.5cm 1cm 0cm 0cm},clip,width=9cm]{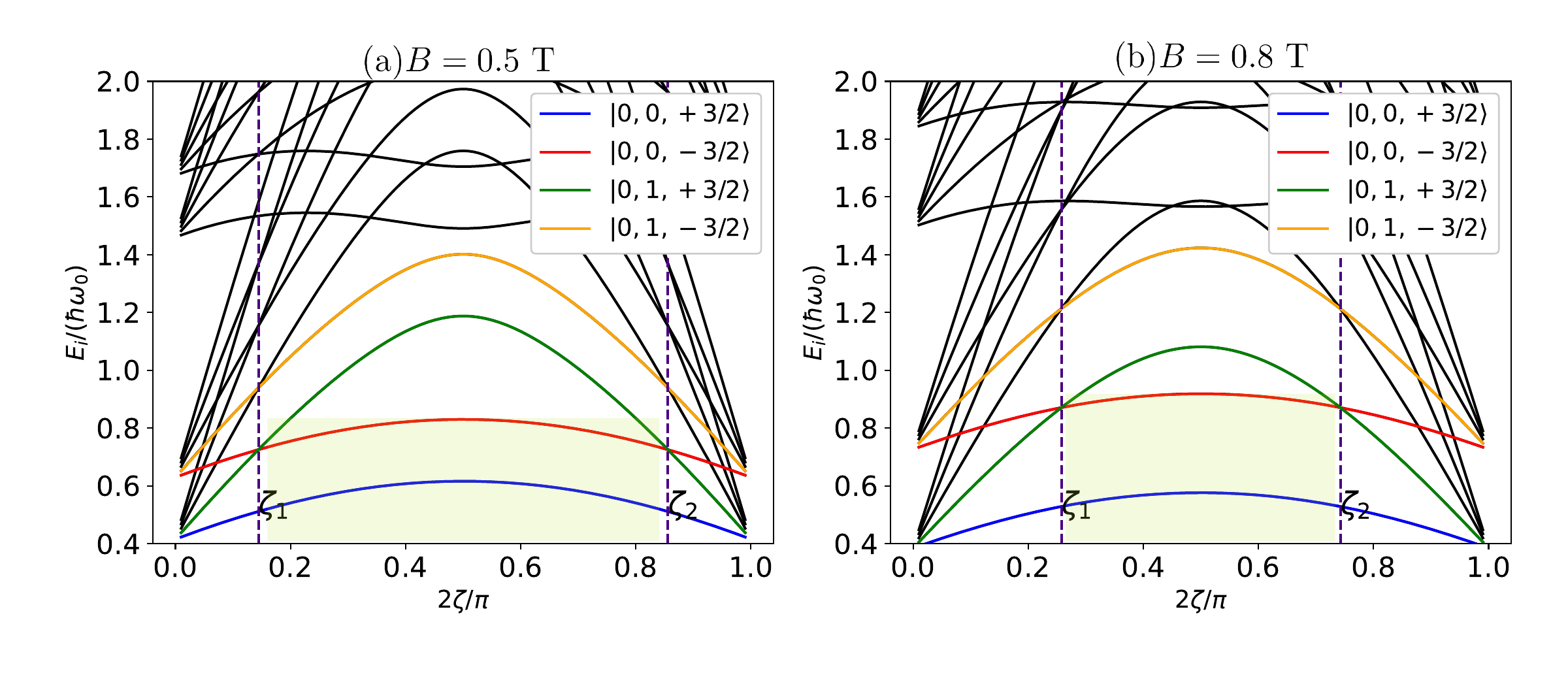}
		\caption{Fock-Darwin energy levels of the anisotropic confinement as a function of $\zeta$ for (a) $B=0.5$ T and (a) $B=0.8$ T. The energy levels $|0,0,-3/2\rangle$ and $|0,1,3/2\rangle$ cross at $\zeta_1$ and $\zeta_2$. The Rabi oscillations are effective for spin-flip operations only for $\zeta_1<\zeta<\zeta_2$.}
		\label{energy}
	\end{figure}

 \begin{figure}[htbp]
		\centering
		\hspace{-0.6cm}\includegraphics[trim={1cm 1cm 0cm 1.3cm},clip,width=9.3cm]{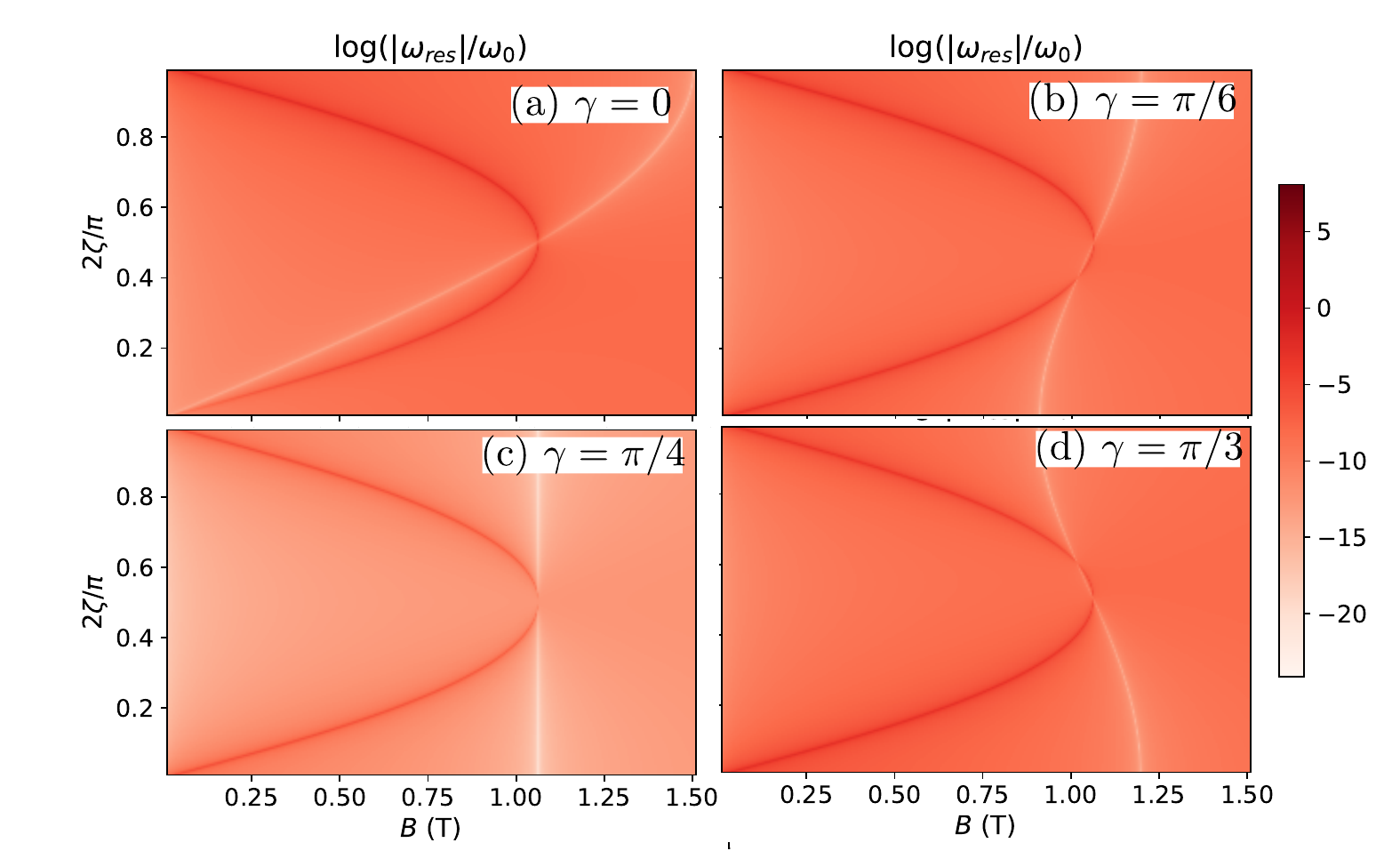}
		\caption{Density plot of the natural logarithm of the resonant Rabi frequency as a function of magnetic field B and $\zeta$ (squeezing angle) for (a) $\gamma=0$ ($x$-polarized), (b)  $\gamma=\pi/6$ (elliptically-polarized), (c) $\gamma=\pi/4$ (circularly-polarized)  and (d) $\gamma=\pi/3$ (same ellipse as (b) but with major and minor axes exchanged).}
		\label{B-zeta}
	\end{figure}
 \begin{figure}[htbp]
		\centering
\includegraphics[trim={0.5cm 0cm 0cm 0cm},clip,width=9cm]{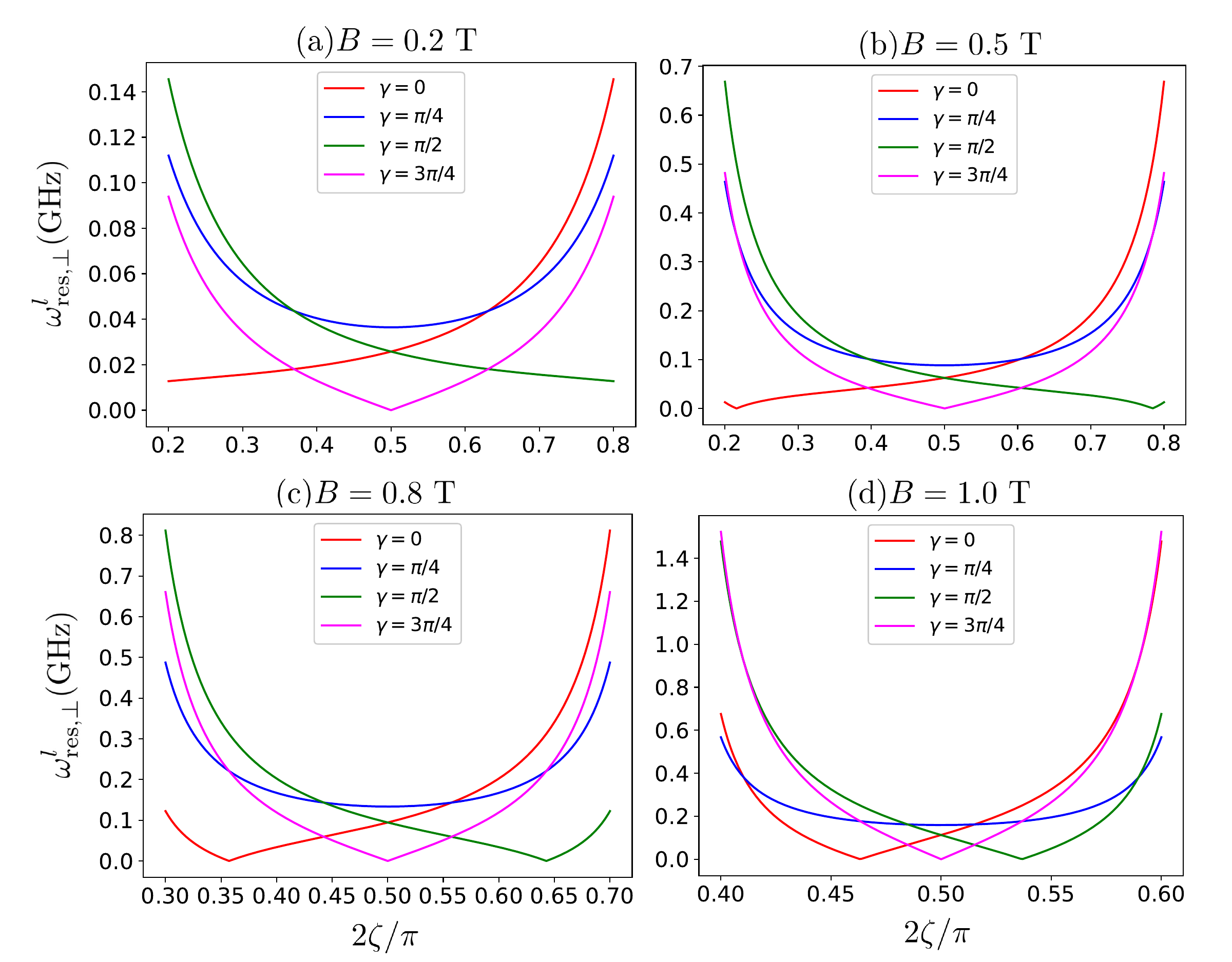}
		\caption{Variation of $|\omega_{\text{res},\perp}^l|$ with squeezing angle $\zeta$ for different polarizations and magnetic field strengths such that $\zeta_1(B)<\zeta<\zeta_2(B)$. }
		\label{wr-vs-zeta-fig}
	\end{figure}
\begin{figure}[htbp]
		\centering
  \vspace{0.2cm}
\includegraphics[trim={0.5cm 0cm 0cm 0cm},clip,width=9cm]{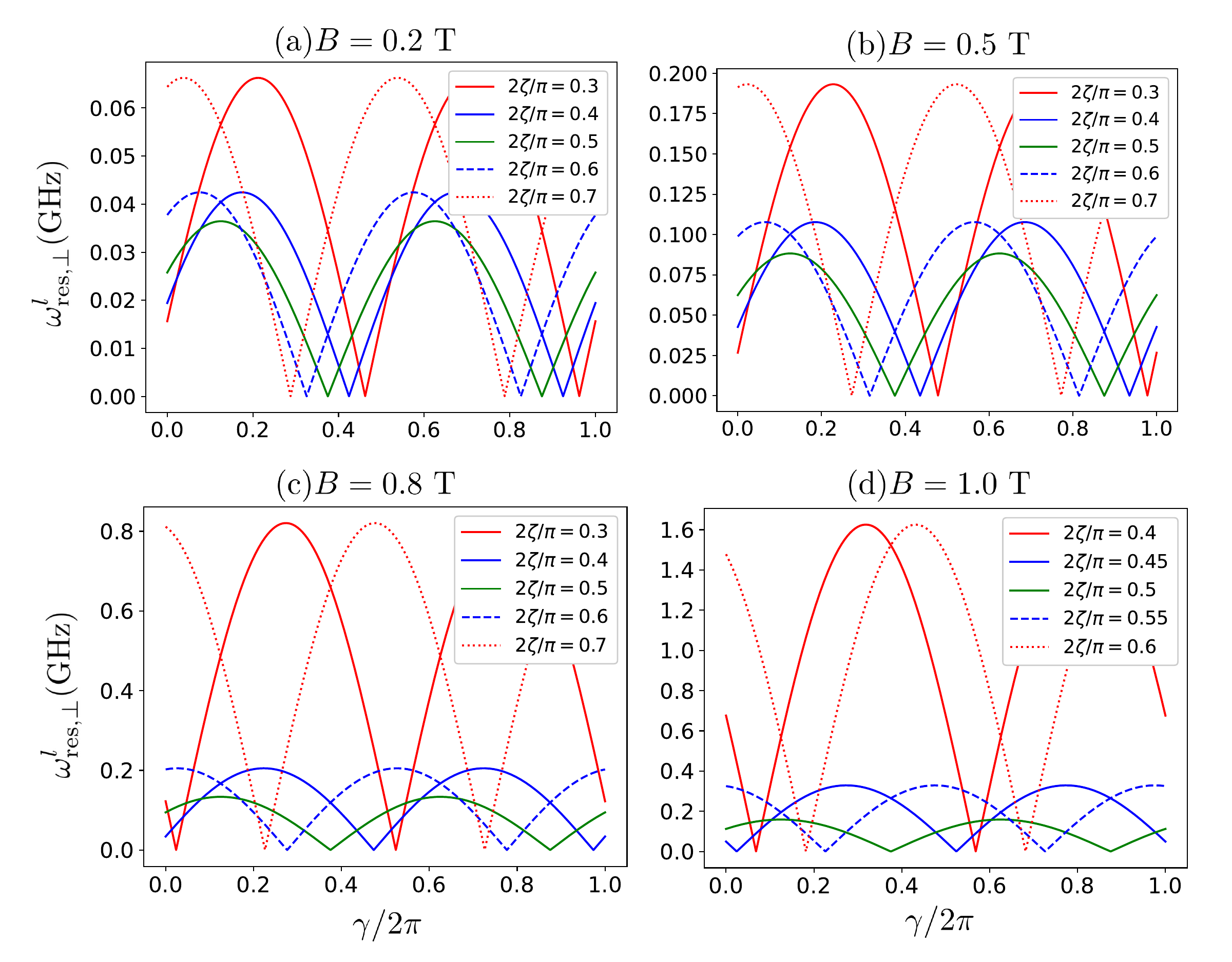}
		\caption{Variation of $|\omega_{\text{res},\perp}^l|$ with polarization angle $\gamma$ for different squeezing angles and magnetic field strengths such that $\zeta_1(B)<\zeta_i<\zeta_2(B)$. }
		\label{wr-vs-gamma-fig}
	\end{figure}

\begin{figure}[htbp]
		\centering
		\hspace{-0.3cm}\includegraphics[trim={0cm 0cm 0cm 0.8cm},clip,width=7cm]{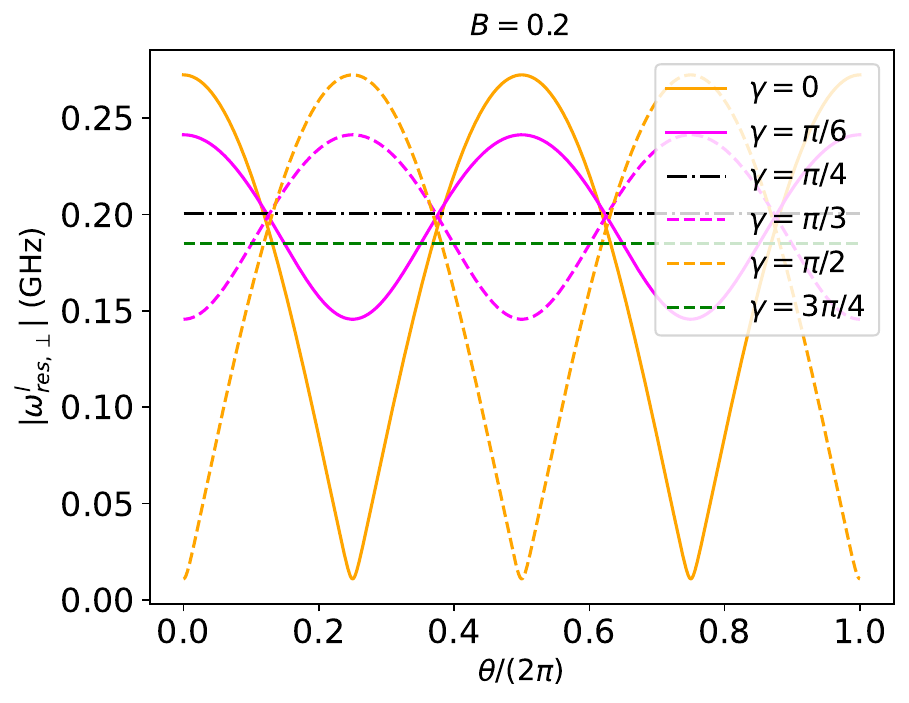}
		\caption{Variation of $|\omega_{\text{res},\perp}^l|$ as a function of $\theta$ (orientation angle) for $\zeta=0.85\pi/2$ and different polarizations at $B=0.2$ T. }
		\label{wr-theta}
	\end{figure}

\section{Results and Discussion}\label{results}

\subsection{Analytical results}\label{results-anal}

Let us parameterize the electric field amplitudes as $E_{0x}=E_0 \cos \gamma$ and $E_{0y}=E_0 \sin \gamma$ where $\gamma$ controls the polarization of the beam. For example, $\gamma=0,\pi/4,\pi/2$ and $3\pi/4$ denote $x$-polarized, left-circular, $y$-polarized and right-circular beams respectively. The driving amplitude $F_0 = |e|\sqrt{E_{0x}^2+E_{0y}^2}=|e| E_0$ is constant with respect to the variation of polarization. This allows us to see purely the polarization effect on the Rabi frequency through the tuning of $\gamma$ without changing the driving strength. Similarly, we can also parameterize the confinement frequencies as $\omega_{x}=\omega_0 \cos\zeta$ and $\omega_{y}=\omega_0 \sin\zeta$  where $\omega_0=\hbar/(m l_0^2)$. Hence, we have $X_0=l_0/\sqrt{\cos \zeta}$ and $Y_0=l_0/\sqrt{\sin \zeta}$. In our calculations, we used $l_0=20$ nm. We label $\gamma$ and $\zeta$ as the `polarization' and `squeezing' angles respectively. The variation of polarization of the beam and contours of the confining potential with $\gamma$ and $\zeta$ respectively are shown in Fig \ref{schematic}. 
 Let us define dimensionless quantities as $\tilde{\omega}_\text{Z}=\omega_\text{Z}/\omega_0$, $\tilde{\omega}_c=\omega_c/\omega_0$, $\tilde{\alpha}_{l}=\alpha_l p_0/(\hbar \omega_0)$, $\tilde{\alpha}_{R/D}^{(+)}=\alpha_{R/D}^{(+)} ~p_0/(\hbar \omega_0)$ and $\tilde{F}_0=F_0 /(p_0\omega_0)$ where $p_0=\sqrt{\hbar m \omega_0}$.  For $l_0=20$ nm and using known values of parameters for Ge/Si quantum wells~\cite{emergence,emergent} i.e. $m\sim0.09~m_e$, $g_\perp\approx15.7$, $g_{||}\approx0.21$,
$\alpha_l=2.01$ meV  A/$\hbar$, we get $\tilde{\alpha}_l=0.0047$, $\tilde{\omega}_c=0.606~B$ $\tilde{\omega}_{Z}=0.428~B$ and $0.00572~ B$ for out-of-plane and in-plane magnetic fields respectively where $B$ is the magnetic field strength in tesla.

\subsubsection{Out-of-plane magnetic field}\label{results-anal-out}

\textbf{Electron-like Rashba SOI: }Figure \ref{gamma-zeta} shows the dependence of $|\omega_{\text{res},\perp}^l|$ on the angles $\gamma$ and $\zeta$. The two dark lines show that the resonant Rabi frequencies sharply peak at two particular values of squeezing angles,  say $\zeta_1$ and $\zeta_2$, which are $B$-dependent and form a pair of complementary angles. This is due to the fact that the energy levels $|0,0,-3/2\rangle$ and $|0,1,3/2\rangle$ cross at $\zeta_{1,2}$ [see Fig. \ref{energy}] in absence of the SOI and are quasidegenerate in presence of it. As a result, some of the $S^{(2)}_{lm}$ given in App. \ref{SWT-app} diverge and the perturbation theory breaks down. Hence, the SWT does not describe the physics correctly at these points.  We shall see in the next section that the Rabi oscillations get heavily distorted close to the lines and completely lose their characteristics at $\zeta_{1,2} (B)$. Hence, the region between but excluding the lines on the $\gamma$-$\zeta$ plane can be termed as the `operating region' for the qubit to perform coherent Rabi oscillations. The fidelity of the operation is lower close these lines.  The lines approach each other with increasing $B$ thereby shrinking the operating region. There also exist curves on which Rabi frequency is vanishingly small. The shape of these curves varies with the magnetic field strength. The range of polarization angle for which we get these diminished frequencies increases with the magnetic field.

Figure \ref{B-zeta} shows the dependence of $|\omega_{\text{res},\perp}^l|$ on magnetic field $B$ and squeezing angle $\zeta$. The peaked values of Rabi frequency trace out curves resembling parabolas on the $B$-$\zeta$ plane. This is also consistent with the existence of a complementary pair ($\zeta_1,\zeta_2$) for a given $B$.  The region enclosed by the curves and the $\zeta$ axis is the operating region for the qubit on the $B$-$\zeta$ plane. The shape of the these curves is independent of the polarization implying that they only depend on the ellipse of the confinement. We can also see curves (light yellow) of diminishing Rabi frequencies whose shapes vary with the polarization. For certain polarizations, a part of the curve lies inside the operating region. For $\gamma=\pi/4$, the curve only touches the region tangentially implying that there is always a resonably high Rabi frequency when the system is driven with left circularly polarized light.

The variation of Rabi frequency with the squeezing angle $\zeta$ is shown in Fig. \ref{wr-vs-zeta-fig} for various polarizations at different magnetic field strengths. The Rabi frequency increases (decreases) with $\zeta$ for $x$-polarized ($y$-polarized) light. This implies that higher Rabi frequency is favored when the ellipse of polarization tends to align with that of the confining potential. With increase in $B$, the Rabi frequency becomes vanishingly small at certain squeezing angles for all but left circularly polarized light ($\gamma=\pi/4$).  
As expected, the variation of the Rabi frequency is symmetric about $\zeta=\pi/4$ i.e circular confinement, for both left and right circularly polarized lights as it should favor squeezing equally along both $x$- and $y$-directions.

The variation of Rabi frequency with the polarization angle $\gamma$ is shown in Fig. \ref{wr-vs-gamma-fig} for various squeezing angles at different magnetic field strengths. For each squeezing angle, the Rabi is $\pi$-periodic in $\gamma$ and diminishes for some $\gamma=\gamma_\zeta(B)$. With an isotropic confinement, the Rabi frequency vanishes for $\gamma=3\pi/4$ at all allowed values of $B$. Using the approach of Ref. \cite{dey}, we find the Rabi frequency for an isotropic dot and elliptical drive to be 
\begin{equation}
    |\omega_{\text{res},\perp}^l|=\frac{2\alpha_l F_0 \omega_\text{Z}}{(\omega_1-\omega_\text{Z})(\omega_2+\omega_\text{Z})}\l(\cos\gamma + \sin\gamma\r)
\end{equation}
where $\omega_{1,2}=\sqrt{\omega^2_0+\omega_c^2/4}\pm\omega_c/2$.
From the above expression, we see that the Rabi frequency vanishes for right circular polarization i.e. $\gamma=3\pi/4$ and $\gamma=7\pi/4$ independent of other parameters.
Maximum Rabi frequency is obtained for values of $\zeta$ close to $\zeta_1$ or $\zeta_2$ i.e. highly squeeezed dots within the operating region.

The variation of the Rabi frequency with orientation angle $\theta$ for $\zeta=0.85\pi/2$ and different polarizations is shown in Fig. \ref{wr-theta}. The Rabi frequency has oscillatory behaviour in $\theta$ with a $\pi$-periodicity for all polarizations except circular. Since the circular polarized radiation is invariant under rotation through $\theta$ (upto a phase), the Rabi frequency is independent of it. Driving with left circular light gives higher Rabi frequency than the right circular one.\\
 
\begin{figure}[htbp]
		\centering
		\hspace{-0.3cm}\includegraphics[trim={0cm 0cm 0cm 0cm},clip,width=9cm]{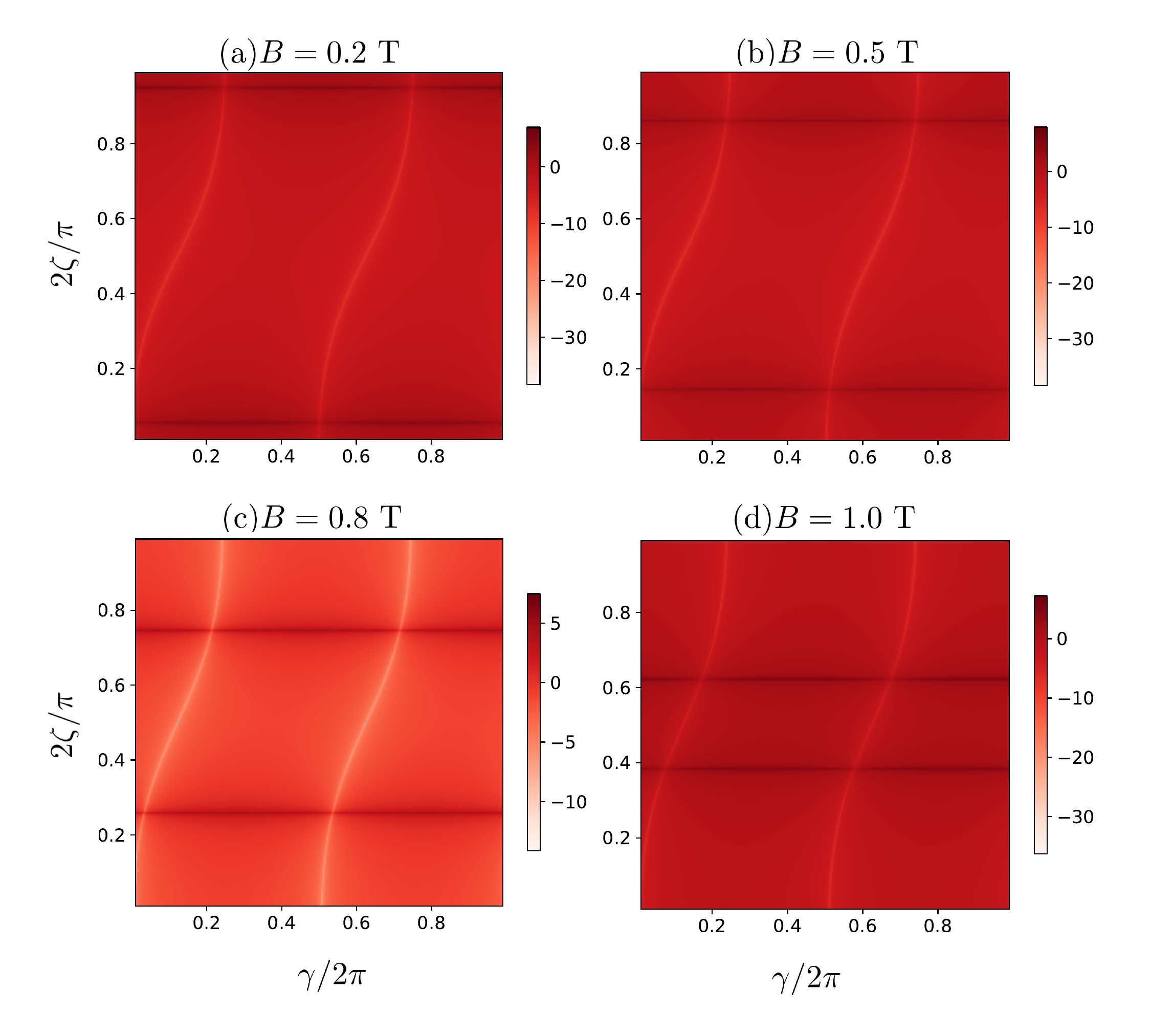}
		\caption{Density plot of the natural logarithm of the $|\omega_{\text{res},\perp}^{(+)}|$ with $\alpha_R\neq0$ and $\alpha_D=0$ as a function of $\gamma$ (polarization angle) and $\zeta$ (squeezing angle) for (a)$B=0.2$ T, (b) $B=0.5$ T, (c)$B=0.8$ T and (d) $B=1$ T. }
		\label{gamma-zeta-holerashba}
	\end{figure}

\textbf{Hole-like Rashba SOI:} Figure \ref{gamma-zeta-holerashba} shows the variation of natural logarithm of $|\omega_{\text{res},\perp}^{(+)}|$ for purely hole-like Rashba SOI i.e. $\alpha_R\neq0 (=\alpha_l)$  and $\alpha_D=0$ with the angles $\gamma$ and $\zeta$. As expected, the operating region which only depends on the ellipse of confinement for a given $B$ is identical to that obtained in the case of electron-like Rashba SOI. However, in contrast to the electron-like Rashba SOI, the curves representing the diminished Rabi frequencies  do not change their shapes with $B$. The curves also have a `horizontally flipped' orientation with respect to that of electron-like Rashba SOI. Unlike the electron-like Rashba SOI, the range of polarization angle for which EDSR is suppressed remains constant with respect to change in the magnetic field. 

\begin{figure}[htbp]
		\centering
		\hspace{-0.3cm}\includegraphics[trim={1cm 0cm 0cm 0cm},clip,width=8.5cm]{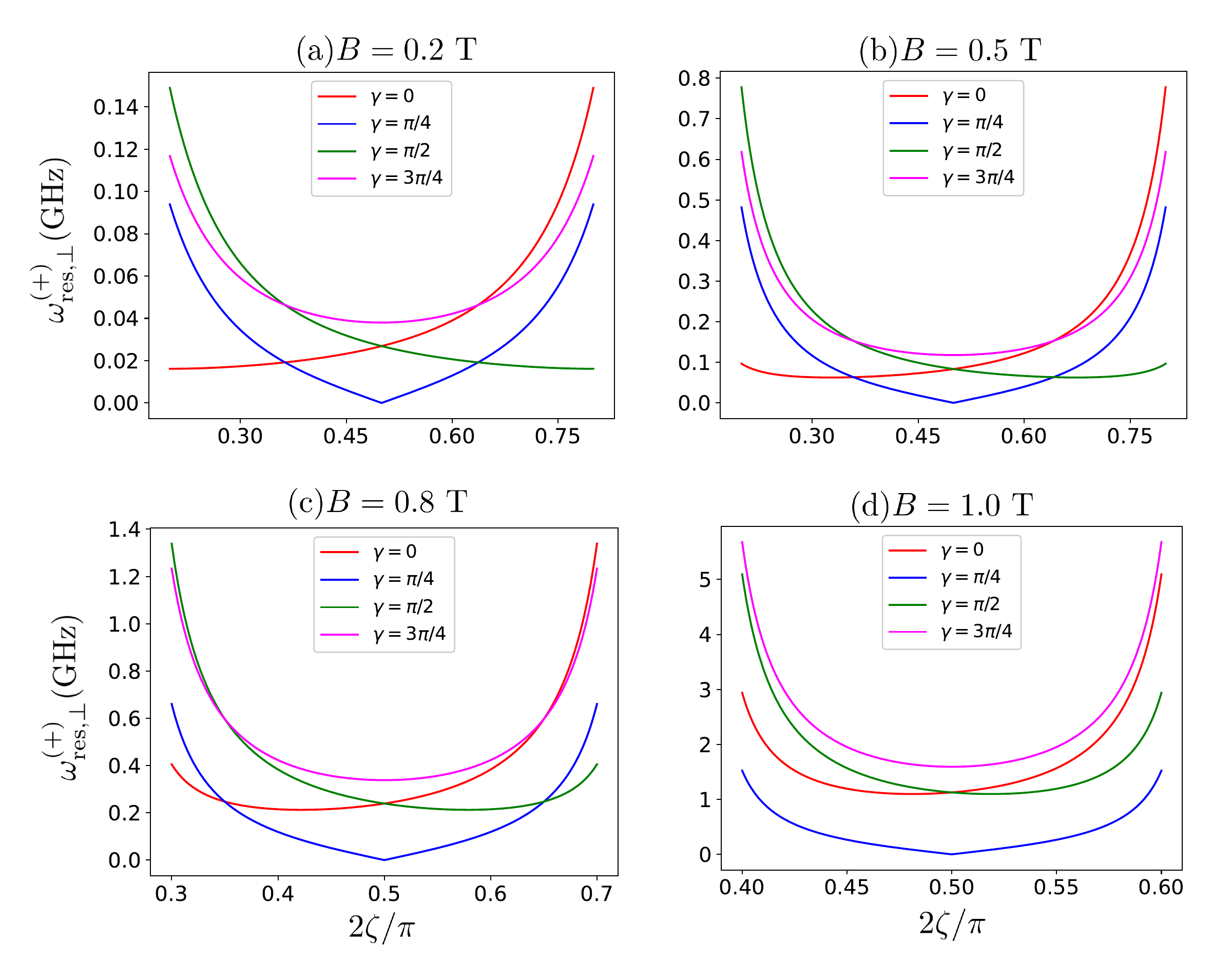}
		\caption{Variation of the $|\omega_{\text{res},\perp}^{(+)}|$ with squeezing angle $\zeta$  for $\alpha_R\neq0$ and $\alpha_D=0$ and different polarization angles $\gamma$ at (a) $B=0.2$ T, (b) $B=0.5$ T, (c)$B=0.8$ T and (d) $B=1$ T. }
		\label{wr-zeta-holerashba}
	\end{figure}

\begin{figure}[htbp]
		\centering
		\hspace{-0.3cm}\includegraphics[trim={1cm 0cm 0cm 0cm},clip,width=9cm]{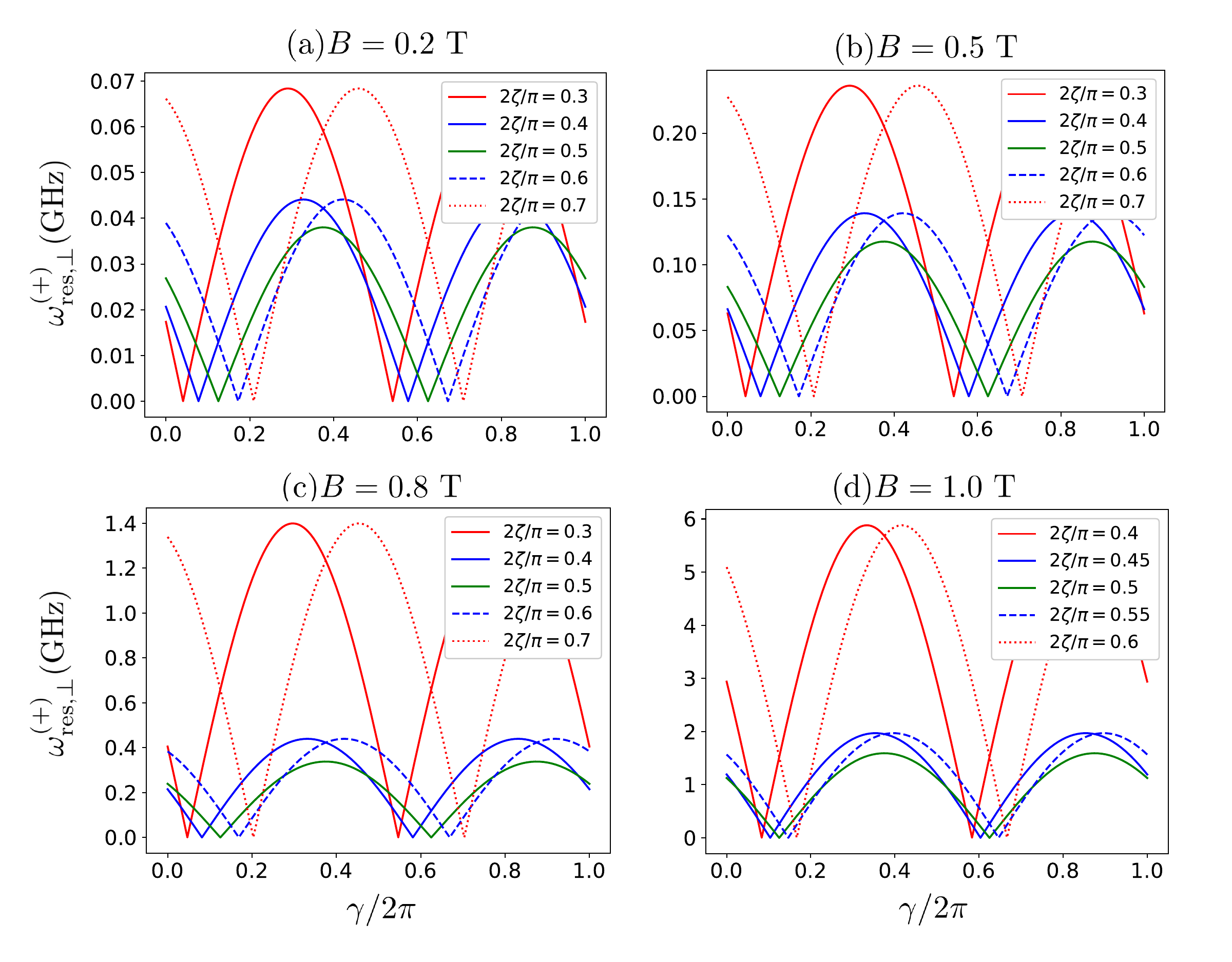}
		\caption{Variation of the $|\omega_{\text{res},\perp}^{(+)}|$ with polarization gangle $\gamma$  for $\alpha_R\neq0$ and $\alpha_D=0$ and different squeezing angles $\zeta$ at (a) $B=0.2$ T, (b) $B=0.5$ T, (c)$B=0.8$ T and (d) $B=1$ T.}
		\label{wr-gamma-holerashba}
	\end{figure}

The variation of $|\omega_{\text{res},\perp}^{(+)}|$ with $\zeta$ is shown in Fig. \ref{wr-zeta-holerashba}. Similar to electron-like Rashba SOI, enhanced Rabi frequencies are observed for higher squeezing when the ellipse of the squeezed configuration is similar to that of the polarization. In contrast to the case of electron-like Rashba SOI, the Rabi frequency vanishes in an isotropic confinement for left circularly polarized light $(\gamma=\pi/4)$ instead of right-circular one. The frequency never diminishes for right circularly polarized light at any squeezed configuration. Hence, the left and right circular polarization switch roles for electron- and hole-like Rashba SOIs. This feature can be used as an experimental probe to decipher the nature of Rashba SOI in heavy holes. Figure \ref{wr-gamma-holerashba} shows the variation of the Rabi frequency with polarization angle $\gamma$. The plots are similar to that of electron-like Rashba SOI except the fact that the point of diminished Rabi frequency for a given $\zeta$ does not change with $B$ in this case. 

\textbf{Hole-like Dresselhaus SOI:} The behaviour of Rabi frequency in presence of purely hole-like Dresselhaus SOI is identical to that of electron-like Rashba SOI. 

\subsubsection{In-plane magnetic field}\label{results-anal-in}

\begin{figure}[htbp]
		\centering
\includegraphics[trim={0cm 0cm 0cm 0cm},clip,width=9cm]{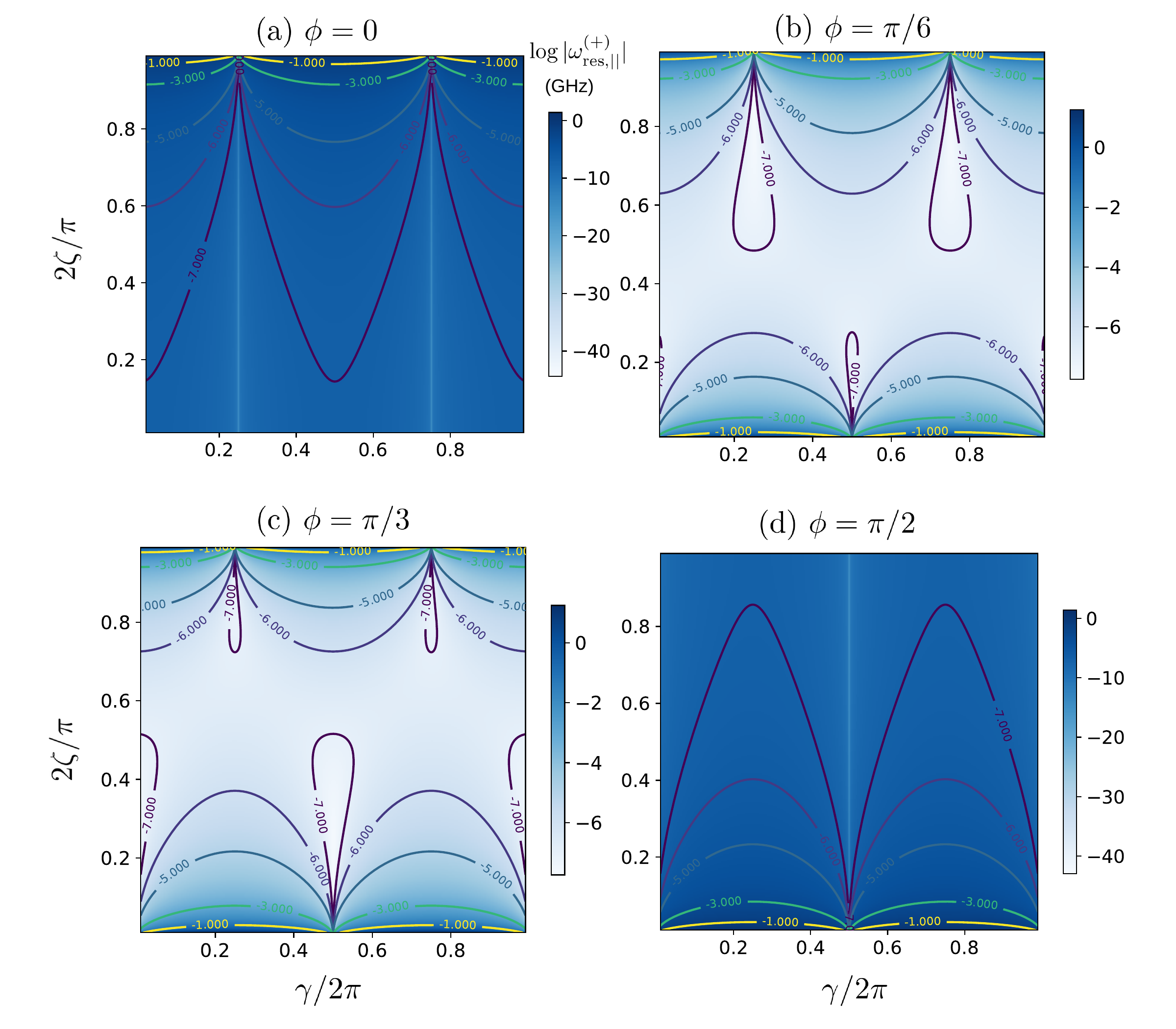}
		\caption{Density plot of $|\omega_{\text{res},||}^l|$ (GHz) for purely Rashba (electron- or hole-like) SOI, given in Eq. (\ref{rabi-big}), as a function of polarization angle ($\gamma$) and squeezing angle ($\zeta$) for different orientations of in-plane magnetic field given by the angle $\phi$.}
		\label{rabi-vs-zeta-gamma-para-Rashba}
	\end{figure}

\textbf{Electron-like Rashba SOI}: The variation of the natural logarithm of $|\omega_{\text{res},||}^l|$ with $\gamma$ and $\zeta$ for electron-like Rashba SOI is shown in Fig. \ref{rabi-vs-zeta-gamma-para-Rashba} for different magnetic field angles $\phi$. Since the $g_{||}\ll g_{\perp}$, we ramp up the magnetic field to 10 T in order to get sufficient Zeeman splitting. Unlike the case of out-of-plane magnetic field, the operating region extends from $\zeta \approx 0$ to $\approx \pi/2$ for all values of $\phi$ and moderate strengths of magnetic field $\sim 10$ T. This is due to the fact that Zeeman splitting is low for in-plane magnetic field allowing for crossing of the energy levels at $\zeta\to0$ and $\zeta\to\pi/2$. The Rabi frequency vanishes at $\gamma=\pi/2, 3\pi/2$ for $\phi=0$ and at $\gamma=0,\pi$\
for $\phi=\pi/2$ as shown by the light vertical lines. This is consistent with the fact the Rabi frequency vanishes when the ${\bf E(t)} \perp {\bf B}$ for Rashba SOI \cite{Dresselhaus}.

\begin{figure}[htbp]
		\centering
\includegraphics[trim={0cm 0cm 0cm 0cm},clip,width=9cm]{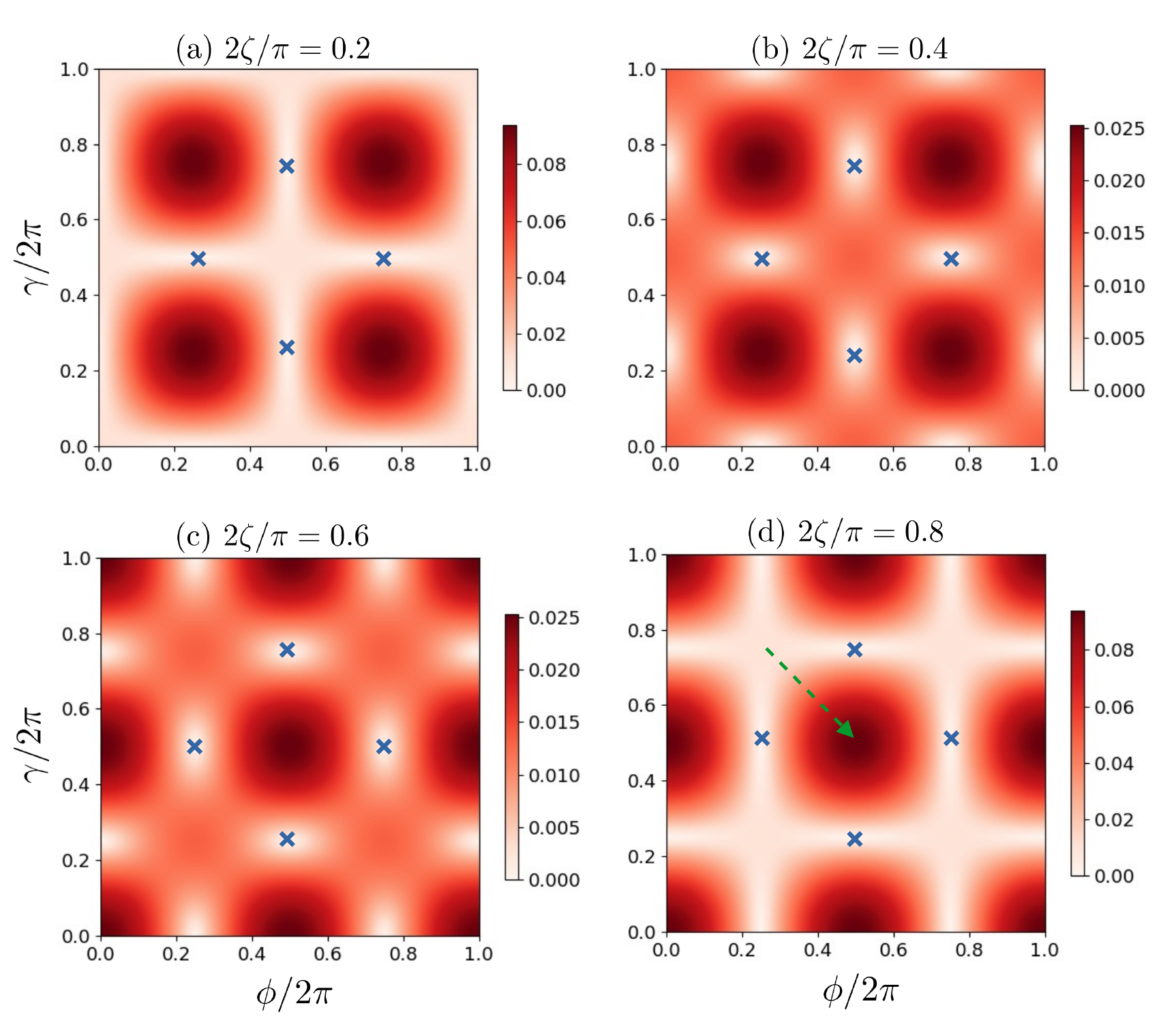}
		\caption{Density plot of $|\omega_{\text{res},||}^l|$ (GHz) for Rashba SOC, given in Eq. (\ref{rabi-big}), as a function of magnetic field angle ($\phi$) and polarization angle ($\gamma$) for different values of squeezing angles $\zeta$. The Rabi frequency vanishes at the crossed points in the plots. The maximas (darkest points on the plots) occur for linearly polarized radiation when ${\bf B}$, ${\bf E} (t)$ and the major axis of the potential contour are aligned in the same direction. The dashed arrow shows the direction along which the maxima shifts as the squeezing angle increases from  $\zeta \to \pi/2-\zeta$. }
		\label{rabi-vs-gamma-phi}
	\end{figure}

In Fig. \ref{rabi-vs-gamma-phi}, we see that the points of vanishing Rabi frequency on the $\phi$-$\gamma$ plane are at $\l[(2n+1)\pi/2,n\pi\r]$ where $n$ is an integer. The variation of the Rabi frequency is $\pi$-periodic in both $\gamma$ and $\phi$. For $\zeta<\pi/4$, the maxima are located at $\l[(2n+1)\pi/2,(2m+1)\pi/2\r]$ while for $\zeta>\pi/4$, the maxima are located at $\l[n \pi, m \pi\r]$ where $n,m$ are integers. The dashed arrow shows the direction along which the maxima shifts as the squeezing angle increases from $\zeta\to\pi/2-\zeta$.\\  

\textbf{Hole-like Rashba and Dresselhaus SOIs}: The behaviour of rabi frequency is identical for electron- and hole-like Rashba SOIs. For hole-like Dresselhaus, the Rabi frequency simply has a phase shift of $\pi/2$ in $\phi$ with respect to that of Rabi driving by Rashba SOI. Consequently, the Rabi frequency vanishes when the ${\bf E(t)} 
|| {\bf B}$ for Dresselhaus SOI \cite{Dresselhaus}.

\begin{figure}[htbp]
		\centering
  \vspace{0.2cm}
\includegraphics[trim={0cm 0cm 0cm 0cm},clip,width=9cm]{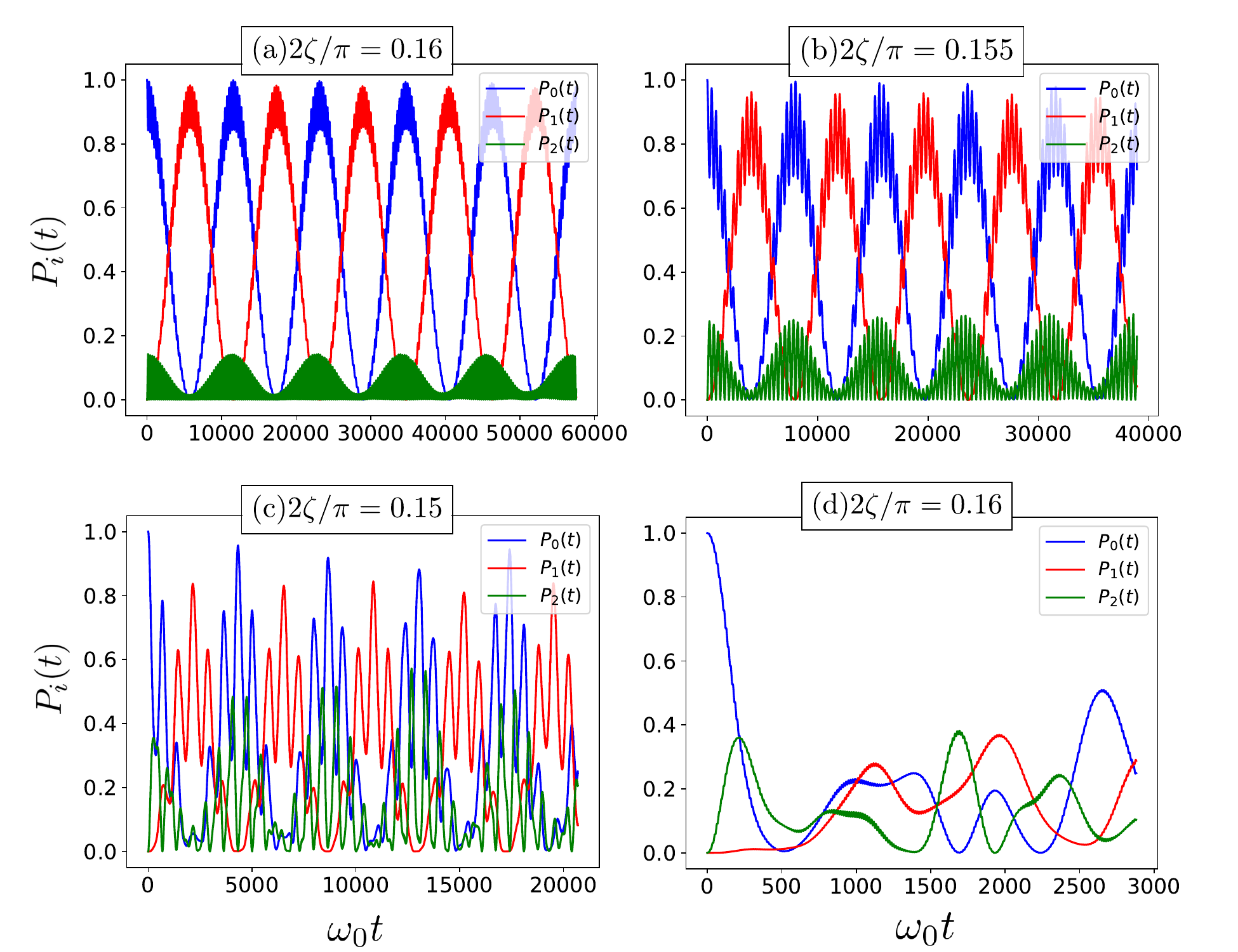}
		\caption{Numerically computed plots (using Floquet theory with 30 Fock-Darwin levels) of the probability oscillations of the levels $|0,0,3/2\rangle$, $|0,0,-3/2\rangle$ and $|0,1,3/2\rangle$ (denoted by $P_0, P_1$ and $P_2$ respectively) vs time for different squeezing angles and under circularly polarized radiation for $B=0.5$ T. The occupation probability of the third level starts to acquire significant values as $\zeta \to \zeta_1$ (plot (d)). Here, $\zeta_1\approx0.144$. In plots (a) and (b), the Rabi frequency is in fairly good agreement with Eq. (\ref{rabiformula}). }
		\label{pt}
	\end{figure}

\subsection{Insights from numerical simulations}\label{results-num}

In this section, we present the numerical results of the time evolution of the qubit for low radiation amplitudes. Since the drive is periodic, we use Floquet theory to compute the time dynamics taking into account 30 energy levels of $H_\text{FD}+H_{\text{Z},\perp}$ or $H_0+H_{\text{Z},||}$ following the methodology given in Ref. \cite{dey}.
The Rabi frequency is found to be in excellent agreement with the numerical values for points within the operating region of the qubit. As $\zeta \to \zeta_1$ or $\zeta_2$ (within the operating region), the oscillations begin to lose their characteristic behaviour and nearly vanish [see Fig. \ref{pt}]. Since energy levels cross at $\zeta_1$ and $\zeta_2$, the effective $2\times2$ Hamiltonian obtained using SW transformation in the $|0,0,\pm3/2\rangle$ block is no longer a good approximation as the interference effects due to the third level become stronger near $\zeta=\zeta_1$ (or $\zeta_2$).  

\subsection{Determination of effective heavy hole mass}

The effective mass of heavy holes can be determined by measuring the Rabi frequency in the presence of an in-plane magnetic field. This can be explained as follows:

From Eq.(\ref{rabi-big}), the Rabi frequencies for ${\bf B}|| \hat{e}_x~ (\phi=0)$ and ${\bf B}|| \hat{e}_y~ (\phi=\pi/2)$ can be written as 
\begin{equation}\label{Bx}
    |\omega_{\text{res}, ||}^l  (0)|=\frac{\alpha_R \omega_\text{Z} F_{0x}}{\hbar(\omega_x^2-\omega_\text{Z}^2)}
\end{equation}
and 
\begin{equation}\label{By}
    |\omega_{\text{res}, ||}^l  (\pi/2)|=\frac{\alpha_R \omega_\text{Z} F_{0y}}{\hbar(\omega_y^2-\omega_\text{Z}^2)},
\end{equation}
respectively. We define a ratio
\begin{equation}
    r=\frac{|\omega_{\text{res}, ||}^l  (0)|}{|\omega_{\text{res}, ||}^l  (\pi/2)|}=\l(\frac{F_{0x}}{F_{0y}}\r)\l( \frac{\omega_y^2-\omega_\text{Z}^2}{\omega_x^2-\omega_\text{Z}^2}\r).
\end{equation}
Defining $f=F_{0x}/F_{0y}$ and using $\omega_x=\hbar/(m X_0^2)$ and $\omega_y=\hbar/(m Y_0^2)$ in the above equation, we get the following expression for $m$ upon some trivial simplifications:
\begin{equation}\label{m}
    m=\frac{\hbar^2}{g_{||} \mu_B B}\sqrt{\l(\frac{1}{r-f}\r)\l(\frac{r}{X_0^4}-\frac{f}{Y_0^4}\r)}.
\end{equation}
Thus, the effective mass can be determined using this expression by experimentally measuring the ratio $r$. Notably, this approach does not require the value of the linear Rashba strength to calculate $m$.

This method for calculating $m$ offers several advantages over conventional techniques such as Shubnikov–de Haas (SdH) oscillations and cyclotron resonance. Unlike SdH oscillations, which rely on resistive measurements and are dissipative in nature, this approach is based on a nearly coherent single-qubit rotation. Determining the hole effective mass at zero doping through SdH oscillations requires fitting the damping of the oscillation amplitude as the temperature increases \cite{wiley2}. In contrast, the Rabi frequency can be obtained through efficient qubit initialization and readout at a fixed cryogenic temperature, eliminating the need for mathematical fitting. Additionally, this method does not require a strong out-of-plane magnetic field to create Landau levels, which is essential for both cyclotron resonance and SdH oscillations. However, our approach also has a limitation. We see from Eq. (\ref{m}) that $m$ is a function of fourth power of the dot's dimensions $X_0$ and $Y_0$. Hence, any error in measuring the values of the dimensions will significantly magnify the error in calculation of $m$.

\section{Conclusion}\label{conc}

We have studied the interplay of squeezing of the confining potential and polarization of the driving electric field on the dynamics of a single hole qubit in a planar germanium quantum dot in presence of $p$-linear SOIs. The squeezing and polarization are parameterized by the angles $\zeta$ and $\gamma$ repectively. We consider two orientations of magnetic field -- in-plane and out-of-plane, which leads to distinct Zeeman couplings owing to the large difference in $g_\perp$ and $g_{||}$ and the anisotropic nature of $g_{||}$. We study the role of electron-like Rashba SOI and hole-like Rashba and Dresselhaus SOI on the Rabi frequencies for each orientation of magnetic field. For an out-of-plane magnetic field, we model the system with the Fock-Darwin Hamiltonian for an anisotropic harmonic potential. We get an operating region on the $\zeta$-$\gamma$ plane bounded by the lines $\zeta=\zeta_1$ and $\zeta=\zeta_2$ within which the qubit can be operated efficiently to obtain high fidelity Rabi oscillations. The oscillations get heavily distorted close to and on these lines. This is attributed to the crossing of higher orbital levels with one of the Zeeman-split levels of the qubit. So, the qubit can no longer be effectively treated as a two-level system. The operating region shrinks with increase in $B$. Higher Rabi frequencies are obtained when the major axes of the ellipses of confinement and polarization are aligned in the same direction. Inside the operating region, curves of highly diminished Rabi frequencies emerge whose shapes of the curves are different for electron- and hole-like Rashba SOIs. The Rabi frequency vanishes for right (left) circular driving in presence of purely electron-like (hole-like) Rashba SOI in a circular confinement. The behaviour of Rabi frequency for hole-like Dresselhaus is identical to that for the electron-like Rashba SOI. The Rabi frequency has a sinusoidal dependence on the orientation angle $\theta$ of the ellipse of polarization. 

For an in-plane magnetic field, the operating regions are approximately $B$-independent and $\zeta_1\approx0$ and $\zeta_2\approx\pi/2$ due to very small $g_{||}$, which corresponds to extremely squeezed configurations. The Rabi frequency vanishes when the driving electric field is linearly polarized with its electric vector perpendicular (parallel) to the static magnetic field in presence of purely electron- or hole-like Rashba (Dresselhaus) SOI. For $\zeta<\pi/4$, the maximum Rabi frequency is obtained when the driving electric field is linearly polarized along $y$-axis with its vector parallel (perpendicular) to the static magnetic field in presence of purely electron- or hole-like Rashba (Dresselhaus) SOI. For $\zeta>\pi/4$, the maximum Rabi frequency is obtained for a similar orientation but with the electric field polarization along the $x$-direction. In both the cases, the maximum value with respect to $\zeta$ occurs for $\zeta \approx 0$ and $\zeta\approx \pi/2$, i.e. highly squeezed configurations. We also demonstrate how the effective mass of heavy holes can be determined by measuring the Rabi frequencies for orthogonal ($x$ and $y$) orientations of the in-plane magnetic field.

Thus, we elucidate the role of squeezing of the confining potential and electric field polarization in the EDSR of a single Ge spin-hole qubit and highlight the operating region of the qubit for distortion-free Rabi oscillations. Although extreme squeezing sharply increases the Rabi frequency, the leakage of higher energy levels into the qubit subspace strongly interferes with the Rabi oscillations, which puts a limitation on the value of the squeezing parameter. Our results highlight the differences in behaviour of the Rabi frequencies for electron/hole-like Rashba and Dresselhaus SOIs in presence of both in-plane and out-of-plane magnetic fields. We have shown that the Rabi frequencies can be significantly enhanced by squeezing the dot (within the perturbative regime) and tuning polarization of the radiation, without the need of increasing the driving and SOI strengths.  In conclusion, our work emphasizes the importance of the geometrical properties of the potential and driving field in EDSR mechanisms. This may offer valuable insights for experimental studies seeking optimal configurations for minimizing the spin-flip times without resorting to stronger electric pulses or SOI strengths, which could increase the decoherence. 

\begin{center}
    {\bf ACKNOWLEDGEMENTS}
\end{center}

This work was funded by the Free state of Bavaria through the ``Munich Quantum valley" as part of a lighthouse project named ``Quantum circuits with spin qubits and hybrid Josephson junctions". We thank Jordi Picó-Cortés and Luca Magazzù for useful discussions.\\

\newpage

\FloatBarrier

\appendix
\section{Schrieffer-Wolff transformation}\label{SWT-app}
Although $V({\bf r},t)$ does not contain spin-mixing terms, it is the combination of SOI and $V({\bf r},t)$ that brings about the desired spin rotations. This can be seen through a Schrieffer-Wolff transformation (SWT) \cite{edsr-loss-orig,phonon2,spin-decay,Fernandez-Fernandez_2023} where we get an effective Rabi Hamiltonian for the spins upon electrical driving by including the effect of SOI perturbatively. In the following, we derive the effective EDSR Hamiltonian for the case of out-of-plane magnetic field and electron-like SOI. Similar approach is to be followed for other cases as well. 

For a small $\alpha_l$, the SWT removes the off-diagonal elements linear in the $\alpha_l$,
\begin{equation}\label{SW}
\begin{aligned}
   H^l_{\text{SW},\perp}&=\e^S (H_\text{FD}+ H_{\text{Z},\perp}+H^l_{\text{SOI},\perp}) \e^{-S}\\
&\approx H_\text{FD}+ H_{\text{Z},\perp}+\frac{1}{2}[S,H^l_{\text{SOI},\perp}].  
\end{aligned}
 \end{equation}
 where $S^\dagger=-S$ and $[H_\text{FD}+ H_{\text{Z},\perp}~,~S]=H_{\text{SOI},\perp}^l$.
Taking the ansatz $S=S^{(1)}\sigma_z+S^{(2)}\sigma_+-S^{(2)^\dagger} \sigma_-$, we get $S^{(1)}=0$ and 
\begin{equation}
    \hat{S}^{(2)}= S^{(2)}_{1a}\hat{a}_1+S^{(2)}_{1b} \hat{a}_1^\dagger+S^{(2)}_{2a} \hat{a}_2+S^{(2)}_{2b} \hat{a}_2^\dagger
\end{equation}
where
\begin{equation}
    S^{(2)}_{1a}=\frac{\alpha_l f_{1-}^{(+)}}{\hbar \omega_1+\hbar \omega_\text{Z}},
\end{equation}
 \begin{equation}
    S^{(2)}_{1b}=\frac{\alpha_l f_{1-}^{(-)}}{\hbar \omega_1-\hbar \omega_\text{Z}},
\end{equation}
 \begin{equation}
    S^{(2)}_{2a}=\frac{-i\alpha_l f_{2+}^{(-)}}{\hbar \omega_2+\hbar \omega_\text{Z}},
\end{equation}
\text{and}
\begin{equation}
    S^{(2)}_{2b}=\frac{-i\alpha_l f_{2+}^{(+)}}{\hbar \omega_2-\hbar \omega_\text{Z}}.
\end{equation}
where $f^{(a)}_{bc}$ are defined in Eqs. (\ref{f_1}) and (\ref{f_2}). Evaluating $[S,H_{\text{SOI},\perp}^l]$ and projecting Eq. (\ref{SW}) into the lowest energy block spanned by the states $|0,0,\pm3/2\rangle$, we get the $2\times 2$ diagonal Hamiltonian
\begin{equation}\label{effective-Bout}
    [H^l_{\text{SW},\perp}]_{2\times2}=\l(\begin{array}{cc}
      E_0 + E_0^{(2)}  & 0 \\
       0  &  E_1 + E_1^{(2)}
    \end{array}\r),
\end{equation}
where $E_{0/1}=\hbar(\omega_1+\omega_2\mp\omega_\text{Z})/2$ are the Zeeman split energies and $E_{0/1}^{(2)}$ are the second order energy corrections in $\alpha_l$ given by
\begin{equation}\label{E02}
    E_0^{(2)}=-\frac{\alpha_l^2}{\hbar}\l[\frac{(f_{1-}^{(+)})^2}{\omega_1+\omega_\text{Z}} + \frac{(f_{2+}^{(-)})^2}{\omega_2+\omega_\text{Z}}\r]
\end{equation}
and 
\begin{equation}\label{E12}
    E_1^{(2)}=-\frac{\alpha_l^2}{\hbar}\l[\frac{(f_{1-}^{(-)})^2}{\omega_1-\omega_\text{Z}} + \frac{(f_{2+}^{(+)})^2}{\omega_2-\omega_\text{Z}}\r].
\end{equation}
Equation (\ref{effective-Bout}) constitutes the effective 2-level Hamiltonian of the spin qubit in this system in absence of an external drive or interaction with environment. 

For a weak electrical driving, the time-dependent SW Hamiltonian can be written upto first order in the driving strength as
\begin{equation}\label{swt}
\begin{aligned}
    H_{\text{SW},\perp}^l(t)&=H^l_{\text{SW},\perp}+\e^S V({\bf r}, t) \e^{-S}\\ & \approx H^l_{\text{SW},\perp}+V({\bf r}, t)+[S,V({\bf r}, t)].
\end{aligned}
\end{equation}
Again, projecting $H_{\text{SW},\perp}^l(t)$ into the lowest energy block, we get a $2\times2$ Hamiltonian as
\begin{equation}\label{SWT22}
\begin{aligned}
    &[H^l_{\text{SW},\perp}]_{2\times2}(t)=\\
    &\l[\begin{array}{cc}
      E_0 + E_0^{(2)}  & \frac{\hbar}{2}(\omega_{\text{res},\perp}^l \e^{i \omega t} +\omega_{\text{off},\perp}^l\e^{-i \omega t})\\
     \frac{\hbar}{2} \{(\omega_{\text{res},\perp}^l)^* \e^{-i \omega t} +(\omega_{\text{off},\perp}^l)^*\e^{i \omega t}\} &  E_1 + E_1^{(2)}
    \end{array}\r].
\end{aligned} 
\end{equation}
Removing the global energy shifts,
we can write the effective EDSR Hamiltonian for the qubit as 
\begin{equation}
\begin{aligned}
    [H^l_{\perp}]_\text{eff}(t)&=-\l(\frac{\hbar \omega_\text{Z}+\Delta^l_\perp}{2}\r)\sigma_z\\
    &+\frac{\hbar }{2} (\omega_{\text{res},\perp}^l \e^{i \omega t} +\omega_{\text{off},\perp}^l\e^{-i \omega t}) \sigma_+ + \text{H.c.}
    \end{aligned}
\end{equation}
where $ \omega_{\text{res},\perp}^l$, $\omega_{\text{off},\perp}^l$ and $\Delta^l_\perp$ are defined in Eqs.  (\ref{rabiformula}), (\ref{off}) and (\ref{Delperp}) respectively. Thus, through SWT, we get an effective Hamiltonian which resembles a Rabi problem with resonant Rabi frequency $|\omega_{\text{res},\perp}^l|$ and resonance condition $\omega=\omega_\text{Z}+\Delta^l_\perp/\hbar$.

\clearpage
\bibliography{bibliography}

\end{document}